\documentclass[12pt]{article}
\usepackage{a4wide}
\usepackage[centertags]{amsmath}
\usepackage{amssymb}
\usepackage[sort&compress,numbers]{natbib}
\usepackage{ifpdf}
\usepackage{graphicx}
\ifpdf
\usepackage[bookmarks]{hyperref}
\else
\usepackage[hypertex]{hyperref}
\fi
\newcommand{\tloc}{\mathcal{T}}
\newcommand{\vp}{\vspace{0.5cm}}
\newcommand{\prn}[1]{\left ( #1 \right )}
\newcommand{\brk}[1]{\left [ #1 \right ]}
\newcommand{\e}{\mathrm{e}}

\newcommand{\preprintno}[3]{\hfill\raisebox{#1}[0cm][0cm]{
\begin{minipage}[t]{#2}\begin{flushright} #3 \end{flushright}\end{minipage}}
\vspace*{-\baselinestretch\baselineskip}}
\newlength{\lingap}
\newcommand{\sgap}{\\[-\lingap]}

\title{Entropy Current in Conformal Hydrodynamics}
\author{R. Loganayagam
\\
\small{\emph{Department of Theoretical Physics, Tata Institute of Fundamental Research,}}\\
\small{\emph{Homi Bhabha Rd, Mumbai 400 005, India}}\sgap
}

\begin{document}

\maketitle

\preprintno{8cm}{6cm}{
   \texttt{TIFR/TH/08-05}
}


\begin{abstract}
In recent work \cite{Bhattacharyya:2007jc,Baier:2007ix}, the energy-momentum tensor for the $\mathcal{N}=4$ SYM fluid was computed up to second derivative terms using holographic methods. The aim of this note is to propose an entropy current (accurate up to second derivative terms) consistent with this energy-momentum tensor and to explicate its relation with the existing theories of relativistic hydrodynamics. In order to achieve this, we first develop a Weyl-covariant formalism  which simplifies the study of conformal hydrodynamics.  This naturally leads us to a proposal for the entropy current of an arbitrary conformal fluid in any spacetime (with $d>3$). In particular, this proposal translates into a definite expression for the entropy flux in the case of $\mathcal{N}=4$ SYM fluid. We conclude this note by comparing the formalism presented here with the conventional Israel-Stewart formalism.
\end{abstract}

\tableofcontents


\section{Introduction}
Many relativistic field theories admit a hydrodynamic description as a low energy approximation.\footnote{In general, hydrodynamics is a valid description of a system when the ratio of mean free path to the length/time scale at consideration (i.e., the Knudsen number) is small and when the system is in local thermal equilibrium to a good approximation.} Relativistic hydrodynamics, in particular, plays an important role in our understanding of various astrophysical phenomena and it appears to be a good description of the physics in the case of heavy-ion collisions at appropriate regimes\footnote{See for example,\cite{Rischke:1995ir,Kolb:2003dz,Shuryak:2003xe,Adams:2005dq,Romatschke:2007mq} and references therein.}.

In this case, the relativistic hydrodynamics relevant for heavy-ion collisions should emerge as an approximation to a strongly coupled field theory - the Quantum Chromodynamics(QCD). One way to develop insight regarding the emergence of hydrodynamic behavior in such a strongly coupled theory is to study the hydrodynamic limit of various other toy models  which are strongly coupled and which somewhat resemble QCD. One such simple model is the $\mathcal{N}=4$ SYM theory which is simpler than QCD because of its super-conformal nature.

Further, $\mathcal{N}=4$ SYM has been conjectured to be dual to II B string theory in the AdS$_5\times $S$^5$ background. This duality is called the AdS/CFT correspondence(See \cite{Aharony:1999ti,Klebanov:2000me,D'Hoker:2002aw} for a review). If we work in a supergravity approximation, the AdS/CFT correspondence relates the thermodynamics of blackholes in the AdS$_5$ background to the thermodynamics of a gauge theory in an appropriate limit. This correspondence has been used extensively to understand $\mathcal{N}=4$ SYM hydrodynamics - including a holographic derivation of the viscosity and more recently, a derivation of  various non-linear response coefficients.

AdS/CFT correspondence is the most well-known example of a more general gauge-gravity duality which conjectures a gravity dual for many gauge theories which need not necessarily be superconformal. Given that we are interested in the low-energy hydrodynamics limit that is dual to AdS gravity, many statements made in this paper can be generalized to hydrodynamics of any such field theory which is dual to general relativity in asymptotically AdS spacetimes.\footnote{There is now a vast literature on hydrodynamic models arising from holography and their applications to heavy ion collisions. A non-exhaustive list of references include \cite{Policastro:2002se,Policastro:2002tn,Herzog:2002fn,Kovtun:2003wp,Kovtun:2004de,Starinets:2005cy,Benincasa:2006fu,Janik:2005zt,Janik:2006ft,Gubser:2006bz,Mas:2006dy,Maeda:2006by,Nakamura:2006ih,Saremi:2006ep,Son:2006em,Lin:2006rf,Lin:2007fa,Liu:2006nn,Liu:2006ug,Heller:2007qt,Kats:2007mq,Kovchegov:2007pq,Myers:2007we,Bhattacharyya:2007jc,Baier:2007ix,Natsuume:2007ty,Kajantie:2008rx} .} In particular, all statements we make about the hydrodynamic description of $\mathcal{N}=4$ SYM will hold true for any four-dimensional conformal field theory with an AdS gravity dual.\footnote{The author wishes to thank Shiraz Minwalla for pointing this out.}

The energy-momentum tensor of $\mathcal{N}=4$ SYM fluid (accurate up to second derivatives of velocity) is now known via holographic methods\cite{Bhattacharyya:2007jc,Baier:2007ix}. In the notation of this paper\footnote{See Appendix(\ref{app:notation}) for a summary of notation used in this paper.} ,
\begin{equation}
\begin{split}
T^{\mu\nu}&= p\ \left(g^{\mu\nu}+4 u^\mu u^\nu \right) +\ \pi^{\mu\nu}\\
\pi^{\mu\nu}&=\quad -2\eta \brk{\sigma^{\mu\nu}-\tau_{\pi}\ u^\lambda \mathcal{D}_\lambda \sigma^{\mu\nu}+\tau_\omega(\omega^\mu{}_\lambda\sigma^{\lambda\nu}+\omega^\nu{}_\lambda\sigma^{\lambda\mu}) }\\
&\quad +\xi_\sigma [\sigma^\mu{}_\lambda\sigma^{\lambda\nu}-\frac{P^{\mu\nu}}{3}\sigma^{\alpha\beta}\sigma_{\alpha\beta}]-\xi_C C^{\mu}{}_{\alpha}{}^{\nu}{}_{\beta}u^\alpha u^\beta\\
\end{split}
\end{equation}
with 
\begin{equation}
\begin{split}
p = \frac{N_c^2}{8\pi^2}(\pi\tloc)^4\  &;\qquad \eta = \frac{N_c^2}{8\pi^2}(\pi\tloc)^3\ ;\\
\qquad \tau_\pi = \frac{ 2- \ln 2 }{2\pi \tloc}\ ;\qquad \tau_\omega = \frac{\ln 2}{2\pi\tloc}\ &;\qquad \xi_\sigma = \xi_C = \frac{4\ \eta}{2\pi\tloc}\ .
\end{split}
\end{equation}
where $p$ is the pressure of the fluid,$\tloc$ its temperature, $u^\mu$ its four-velocity and $\eta$ its shear viscosity. The second equation is the constitutive relation that relates the visco-elastic stress $\pi^{\mu\nu}$ to the shear strain rate $\sigma^{\mu\nu}$ and vorticity $\omega^{\mu\nu}$.  $\tau_{\pi},\tau_\omega,\xi_\sigma$ and $\xi_C$ are the non-linear response coefficients.

In this paper, we propose an entropy current consistent with the energy-momentum tensor above -
\begin{equation}
\begin{split}
J^\lambda_s &=4\pi\eta \left[u^\lambda -\frac{\left[(\ln {2}) \sigma^{\mu\nu} \sigma_{\mu\nu}+\omega^{\mu\nu} \omega_{\mu\nu}\right]u^\lambda+2 u_\mu(\mathcal{G}^{\mu\lambda}+\mathcal{F}^{\mu\lambda})+6\mathcal{D}_\nu\omega^{\lambda\nu}}{8(\pi\tloc)^2}\right].\\
\qquad \text{with}& \qquad  \tloc \mathcal{D}_\mu J^\mu_S =  2\eta \sigma^{\mu\nu} \sigma_{\mu\nu} \geq 0\\
\end{split}
\end{equation}
Note that the above expression,  reduces in the appropriate limit to the holographic result $ J^\lambda_s =4\pi\eta u^\lambda $ of Kovtun, Son and Starinets\cite{Kovtun:2004de}.

The plan of the paper is as follows -  In \S\ref{sec:confobs}, we introduce a manifestly Weyl-covariant derivative especially suited to the study of conformal fluids and list the various conformal observables that occur in fluid mechanics. Since, we are interested in conformal fluids on arbitrary spacetimes, in section \S\ref{sec:curv} we describe in some detail the various curvature related observables that occur in conformal hydrodynamics. This is followed by the section \S\ref{sec:confhydro}, where the equations of fluid mechanics are formulated in a conformally covariant way.  We end \S\ref{sec:confhydro} by writing down the derivative expansion for a conformal fluid exact up to second derivative terms.

Next, we proceed in section \S\ref{sec:entropy} to find a derivative expansion of the local entropy current for a conformal fluid which obeys the second law of thermodynamics. We make a proposal for the entropy current of a conformal fluid living in arbitrary spacetimes (with $d>3$). Next,in section \S\ref{sec:adsfluid}, we turn to the specific case of $\mathcal{N}=4$ SYM and find the corresponding expression for the entropy flux.

This is followed by the section \S\ref{sec:relhyd} where we compare  the method adopted in this paper with the  existing theories of relativistic hydrodynamics. In the final section, we discuss
future directions and conclude. In appendix (\ref{sec:identity}) , we prove some very useful identities that were used in the body of the paper. This is followed by appendix (\ref{sec:cfnofR}) where we discuss the various terms that can in principle occur in the energy-momentum tensor of a conformal fluid. Finally, appendix (\ref{app:notation}) has a summary of notation used in this paper.

\section{Conformal Observables in Hydrodynamics}\label{sec:confobs}
In the following section, we first introduce a manifestly Weyl-covariant formalism which is especially suited to the study of conformal fluids. This is followed by a brief discussion on the various conformal observables in fluid mechanics. 

Consider a conformal fluid in $d>3$ dimensions. We seek the Weyl transformations of various observables of such a fluid. To this end, consider a conformal transformation which replaces the old metric $g_{\mu\nu}$ with $\tilde{g}_{\mu\nu}$ given by
\begin{equation}\label{g:eq}
  g_{\mu\nu} = \e^{2\phi}\tilde{g}_{\mu\nu}; \qquad
 g^{\mu\nu} = \e^{-2\phi}\tilde{g}^{\mu\nu}
\end{equation}
The Christoffel symbols transform as(See, for example, appendix (D) of \cite{1984ucp..book.....W})
\begin{equation}\label{Gamma:eq}
\begin{split}
\Gamma_{\lambda\mu}{}^{\nu}&= \widetilde{\Gamma}_{\lambda\mu}{}^{\nu} + \delta^{\nu}_{\lambda}\partial_{\mu}\phi+ \delta^{\nu}_{\mu}\partial_{\lambda}\phi-
 \tilde{g}_{\lambda\mu}\tilde{g}^{\nu\sigma}\partial_{\sigma}\phi
\\
\end{split}
\end{equation}

Let $u^{\mu}$ be the four-velocity describing the fluid motion. Using $g_{\mu\nu}u^{\mu}u^{\nu}=\tilde{g}_{\mu\nu}\tilde{u}^{\mu}\tilde{u}^{\nu}=-1$, we get $u^\mu=\e^{-\phi}\tilde{u}^\mu$. It follows that the projection tensor transforms as $P^{\mu\nu}=g^{\mu\nu}+u^{\mu}u^{\nu}=\e^{-2\phi}\widetilde{P}^{\mu\nu}$. The transformation of the covariant derivative of $u^{\mu}$ is given by
\begin{equation}\label{gradu:eq}
\begin{split}
\nabla_{\mu}u^{\nu}&=\partial_{\mu}u^{\nu}+\Gamma_{\mu\lambda}{}^{\nu}u^{\lambda}\\
&=\e^{-\phi}\brk{\widetilde{\nabla}_{\mu}\tilde{u}^{\nu}+\delta^{\nu}_{\mu}\tilde{u}^{\sigma}\partial_{\sigma}\phi- \tilde{g}_{\mu\lambda}\tilde{u}^{\lambda}\tilde{g}^{\nu\sigma}\partial_{\sigma}\phi}
\end{split}
\end{equation}

The above equation can be used to derive the transformation of various related quantities
\begin{equation}\label{thetaaA:eq}
\begin{split}
  \vartheta &\equiv \nabla_{\mu}u^{\mu}
  =\e^{-\phi}\brk{\tilde{\vartheta}+(d-1)\tilde{u}^{\sigma}\partial_{\sigma}\phi},\\
 a^{\nu} &\equiv u^{\mu}\nabla_{\mu}u^{\nu}
  =\e^{-2\phi}\brk{\tilde{a}^{\nu}+\widetilde{P}^{\nu\sigma}\partial_{\sigma}\phi},\\
\mathcal{A}_{\nu} &\equiv a_{\nu}-  \frac{\vartheta}{d-1} u_{\nu} = \tilde{\mathcal{A}}_{\nu} + \partial_{\nu}\phi .
\end{split}
\end{equation}

We define a Weyl covariant derivative \footnote{More precisely, what we are doing here is to use the additional  mathematical structure provided by a fluid background (namely a unit time-like vector field with conformal weight $w=1$) to define what is known as a \textit{Weyl connection} over $(\mathcal{M},\mathcal{C})$ where $\mathcal{M}$ is the spacetime manifold with the conformal class of metrics $\mathcal{C}$ . A torsionless connection $\nabla^{weyl}$ is called a Weyl connection(see for example, \cite{1992JMP....33.2633H} and references therein)  if for every metric in the conformal class $\mathcal{C}$ there exists a one form $\mathcal{A}_\mu$ such that $\nabla^{weyl}_\mu g_{\nu\lambda}= 2 \mathcal{A}_\mu g_{\nu\lambda}$ . Having a fluid over the manifold provides us a natural one form $\mathcal{A}_\mu$ (see below), which can in turn be used to define a Weyl connection. The `prolonged' covariant derivative $\mathcal{D}$ that we use in this paper is related to this Weyl connection via the relation $\mathcal{D}_\mu = \nabla^{weyl}_\mu + w\mathcal{A}_\mu $ . In terms of this covariant derivative, the condition for Weyl connection is just the statement of metric compatibility($ \mathcal{D}_{\lambda}g_{\mu\nu} =0 $) and the one-form $\mathcal{A}_\mu$ is uniquely determined  by requiring that the covariant derivative of $u^\mu$ be transverse ($u^\lambda\mathcal{D}_\lambda u^\mu = 0 $) and traceless ($\mathcal{D}_\lambda u^\lambda =0 $).} $\mathcal{D}$ such that,  if a tensorial quantity $Q^{\mu\ldots}_{\nu\ldots}$ obeys $Q^{\mu\ldots}_{\nu\ldots}= e^{-w\phi}\widetilde{Q}^{\mu\ldots}_{\nu\ldots}$ , then $\mathcal{D}_\lambda\ Q^{\mu\ldots}_{\nu\ldots}= e^{-w\phi} \widetilde{\mathcal{D}}_\lambda\widetilde{Q}^{\mu\ldots}_{\nu\ldots}$ where
\begin{equation}\label{D:eq}
\begin{split}
\mathcal{D}_\lambda\ Q^{\mu\ldots}_{\nu\ldots} &\equiv \nabla_\lambda\ Q^{\mu\ldots}_{\nu\ldots} + w\  \mathcal{A}_{\lambda} Q^{\mu\ldots}_{\nu\ldots} \\ &+\brk{{g}_{\lambda\alpha}\mathcal{A}^{\mu} - \delta^{\mu}_{\lambda}\mathcal{A}_\alpha  - \delta^{\mu}_{\alpha}\mathcal{A}_{\lambda}} Q^{\alpha\ldots}_{\nu\ldots} + \ldots\\
&-\brk{{g}_{\lambda\nu}\mathcal{A}^{\alpha} - \delta^{\alpha}_{\lambda}\mathcal{A}_\nu  - \delta^{\alpha}_{\nu}\mathcal{A}_{\lambda}} Q^{\mu\ldots}_{\alpha\ldots} - \ldots
\end{split}
\end{equation}
Note that the above covariant derivative is metric compatible ($\mathcal{D}_{\lambda}g_{\mu\nu} =0 $).

Using the Weyl covariant derivative, the fluid mechanics can be cast into a manifestly conformal language. In order to make contact with the conventional fluid dynamics, we give below some commonly occurring observables in both the notations - the advantages of the manifestly conformal notation is self-evident.
\begin{equation}\label{Du:eq}
\begin{split}
\mathcal{D}_{\mu}u^{\nu} &=  \nabla_{\mu}u^{\nu} + u_{\mu}a^{\nu} -  \frac{\vartheta}{d-1} P_{\mu}{}^{\nu}  = \sigma_{\mu}{}^{\nu} + \omega_{\mu}{}^{\nu} =\e^{-\phi}\widetilde{\mathcal{D}}_{\mu}\tilde{u}^{\nu} , \\
  \sigma^{\mu\nu} &\equiv \frac{1}{2} \prn{P^{\mu\lambda} \nabla_\lambda u^\nu
                   + P^{\nu\lambda} \nabla_\lambda u^\mu}
                   - \frac{1}{d-1} \vartheta P^{\mu\nu}
= \frac{1}{2} \prn{\mathcal{D}^{\mu} u^\nu + \mathcal{D}^{\nu} u^\mu}
                   = \e^{-3\phi} \tilde{\sigma}^{\mu\nu},\\
  \omega^{\mu\nu} &\equiv \frac{1}{2} \prn{P^{\mu\lambda} \nabla_\lambda u^\nu
                  - P^{\nu\lambda} \nabla_\lambda u^\mu}
=  \frac{1}{2} \prn{\mathcal{D}^{\mu} u^\nu - \mathcal{D}^{\nu} u^\mu} = \e^{-3\phi} \widetilde{\omega}^{\mu\nu}.
\end{split}
\end{equation}

In order to study fluid dynamics up to second derivative terms, we will need the expressions involving second derivatives of fluid velocity. 
\begin{equation}\label{DDu:eq}
\begin{split}
\mathcal{D}_{\mu}\mathcal{D}_{\nu} u^\lambda &= \mathcal{D}_{\mu} {\sigma}_{\nu}{}^{\lambda} + \mathcal{D}_{\mu}{\omega}_{\nu}{}^{\lambda} = \e^{-\phi}\widetilde{\mathcal{D}}_{\mu}\widetilde{\mathcal{D}}_{\nu}\tilde{u}^{\lambda}\\
\mathcal{D}_{\lambda} {\sigma}_{\mu\nu} &= \nabla_{\lambda}{\sigma}_{\mu\nu} + \mathcal{A}_{\lambda}{\sigma}_{\mu\nu}+\mathcal{A}_{\mu}{\sigma}_{\lambda\nu}+\mathcal{A}_{\nu}{\sigma}_{\mu\lambda} - g_{\mu\lambda}\mathcal{A}^{\alpha}{\sigma}_{\alpha\nu} - g_{\nu\lambda}\mathcal{A}^{\alpha}{\sigma}_{\mu\alpha}= \e^{\phi} \widetilde{\mathcal{D}}_{\lambda} \tilde{\sigma}_{\mu\nu} \\
\mathcal{D}_{\lambda} {\omega}_{\mu\nu} &= \nabla_{\lambda}{\omega}_{\mu\nu} + \mathcal{A}_{\lambda}{\omega}_{\mu\nu}+\mathcal{A}_{\mu}{\omega}_{\lambda\nu}+\mathcal{A}_{\nu}{\omega}_{\mu\lambda} - g_{\mu\lambda}\mathcal{A}^{\alpha}{\omega}_{\alpha\nu}- g_{\nu\lambda}\mathcal{A}^{\alpha}{\omega}_{\mu\alpha}=\e^{\phi} \widetilde{\mathcal{D}}_{\lambda} \tilde{\omega}_{\mu\nu}
\end{split}
\end{equation}

Apart from the fluid velocity $u^\mu$ introduced above, a conformal fluid is characterized by its temperature $\tloc$ and various chemical potentials $\mu_i$ associated with different conserved charges(where $i=1,\ldots,k$ denotes the various charge currents). Under the AdS/CFT correspondence, these thermodynamic quantities can be directly related to the thermodynamic properties of black holes in the AdS backgrounds.

The Weyl transformation of the temperature and the chemical potentials can be written as $\tloc=\e^{-\phi}\widetilde{\tloc}$ and $\mu_i=\e^{-\phi}\tilde{\mu}_i$ . Further, we can define $\nu_i={\mu_i}/{\tloc}=\tilde{\nu_i}$ . It is straightforward to write down the conformal observables involving no more than second derivatives of the temperature and the chemical potentials.
\begin{equation}\label{DnuDT:eq}
\begin{split}
\mathcal{D}_\mu\nu_i &= \nabla_\mu\nu_i = \widetilde{\mathcal{D}}_\mu\tilde{\nu}_i  ,\qquad \mathcal{D}_\mu \tloc = (\nabla_\mu + \mathcal{A}_\mu)\tloc = \e^{-\phi} \widetilde{\mathcal{D}}_\mu \widetilde{\tloc} \\
\mathcal{D}_\lambda\mathcal{D}_\sigma\nu_i &= \nabla_\lambda\nabla_\sigma\nu_i + \mathcal{A}_\lambda\nabla_\sigma\nu_i + \mathcal{A}_\sigma\nabla_\lambda\nu_i - g_{\lambda\sigma}\mathcal{A}^\alpha\nabla_\alpha\nu_i = \widetilde{\mathcal{D}}_\lambda\widetilde{\mathcal{D}}_\sigma\tilde{\nu}_i \\
\mathcal{D}_\lambda\mathcal{D}_\sigma\tloc &= \nabla_\lambda\nabla_\sigma\tloc + 2 \mathcal{A}_\lambda\nabla_\sigma\tloc + 2 \mathcal{A}_\sigma\nabla_\lambda\tloc - g_{\lambda\sigma}\mathcal{A}^\alpha\nabla_\alpha\tloc \\
&\qquad + \tloc\brk{\nabla_\lambda\mathcal{A}_\sigma + 3 \mathcal{A}_\lambda\mathcal{A}_\sigma - g_{\lambda\sigma}\mathcal{A}^\alpha\mathcal{A}_\alpha} = \e^{-\phi} \widetilde{\mathcal{D}}_\lambda\widetilde{\mathcal{D}}_\sigma\widetilde{\tloc}  
\end{split}
\end{equation}

Fortunately, we rarely have to deal with the above quantities in their entirety. Often, only specific projections of the above quantities are required. We list below some common fluid mechanical observables which involve second derivative of the fluid velocity -
\begin{equation}\label{DDuProj:eq}
\begin{split}
\mathcal{D}^\lambda {\sigma}_{\mu\lambda} &= \left(\nabla^\lambda-(d-1)\mathcal{A}^\lambda\right){\sigma}_{\mu\lambda} = \e^{\phi} \widetilde{\mathcal{D}}^\lambda \tilde{\sigma}_{\mu\lambda}\\
\mathcal{D}^\lambda {\omega}_{\mu\lambda} &= \left(\nabla^\lambda-(d-3)\mathcal{A}^\lambda\right){\omega}_{\mu\lambda}= \e^{\phi} \widetilde{\mathcal{D}}^\lambda \tilde{\omega}_{\mu\lambda} \\
u^\lambda\mathcal{D}_{\lambda} {\sigma}_{\mu\nu} &= u^\lambda\nabla_{\lambda}{\sigma}_{\mu\nu} + \frac{\vartheta}{d-1}{\sigma}_{\mu\nu} - u_{\mu}\mathcal{A}^{\alpha}{\sigma}_{\alpha\nu} - u_{\nu}\mathcal{A}^{\alpha}{\sigma}_{\alpha\mu} = \tilde{u}^\lambda\widetilde{\mathcal{D}}_{\lambda} \tilde{\sigma}_{\mu\nu} \\
&=P_{\mu}{}^{\alpha}P_\nu{}^{\beta}u^{\lambda}\mathcal{D}_{\lambda} {\sigma}_{\alpha\beta}= P_{\mu}{}^{\alpha}P_\nu{}^{\beta}u^{\lambda}\nabla_{\lambda}{\sigma}_{\alpha\beta} + \frac{\vartheta}{d-1}{\sigma}_{\mu\nu} \\
u^\lambda\mathcal{D}_{\lambda} {\omega}_{\mu\nu} &= u^\lambda\nabla_{\lambda}{\omega}_{\mu\nu} + \frac{\vartheta}{d-1}{\omega}_{\mu\nu} - u_{\mu}\mathcal{A}^{\alpha}{\omega}_{\alpha\nu} + u_{\nu}\mathcal{A}^{\alpha}{\omega}_{\alpha\mu} = \tilde{u}^\lambda\widetilde{\mathcal{D}}_{\lambda} \tilde{\omega}_{\mu\nu} \\
&=P_{\mu}{}^{\alpha}P_\nu{}^{\beta}u^{\lambda}\mathcal{D}_{\lambda} {\omega}_{\alpha\beta}= P_{\mu}{}^{\alpha}P_\nu{}^{\beta}u^{\lambda}\nabla_{\lambda}{\omega}_{\alpha\beta} + \frac{\vartheta}{d-1}{\omega}_{\mu\nu} \\
u^\mu\mathcal{D}_{\lambda} {\sigma}_{\mu\nu} &= u^\mu\nabla_{\lambda}{\sigma}_{\mu\nu} + \frac{\vartheta}{d-1}{\sigma}_{\lambda\nu} - u_{\lambda}\mathcal{A}^{\alpha}{\sigma}_{\alpha\nu} = \tilde{u}^\mu\widetilde{\mathcal{D}}_{\lambda} \tilde{\sigma}_{\mu\nu} \\
&= -(\mathcal{D}_{\lambda}u^\mu) {\sigma}_{\mu\nu} = -\sigma_{\lambda}{}^\mu {\sigma}_{\mu\nu} -\omega_{\lambda}{}^\mu {\sigma}_{\mu\nu}  \\
u^\mu\mathcal{D}_{\lambda} {\omega}_{\mu\nu} &= u^\mu\nabla_{\lambda}{\omega}_{\mu\nu} - \frac{\vartheta}{d-1}{\omega}_{\lambda\nu} - u_{\lambda}\mathcal{A}^{\alpha}{\omega}_{\alpha\nu} = \tilde{u}^\mu\widetilde{\mathcal{D}}_{\lambda} \tilde{\omega}_{\mu\nu} \\
&= -(\mathcal{D}_{\lambda}u^\mu) {\omega}_{\mu\nu} = -\sigma_{\lambda}{}^\mu {\omega}_{\mu\nu} -\omega_{\lambda}{}^\mu {\omega}_{\mu\nu}  \\
\end{split}
\end{equation}

All observables in conformal hydrodynamics (that is accurate up to second derivative terms) can be written in terms of the following quantities -
\begin{equation}\label{confobs:eq}
\begin{split}
&\nu_i,\ \tloc,\ u^{\mu},\ g_{\mu\nu},\ \epsilon^{\mu\nu\ldots\sigma} \\
&\mathcal{D}_\mu\nu_i,\ \mathcal{D}_\mu \tloc,\ \sigma_{\mu\nu},\ \omega_{\mu\nu}, \\
&\mathcal{D}_\lambda\mathcal{D}_\sigma\nu_i,\ \mathcal{D}_\lambda\mathcal{D}_\sigma\tloc,\  \mathcal{F}_{\mu\nu} =\nabla_\mu 
\mathcal{A}_\nu-\nabla_\nu \mathcal{A}_\mu ,\  \mathcal{D}_\lambda {\sigma}_{\mu\nu} ,\  \mathcal{D}_\lambda {\omega}_{\mu\nu},\\
&\mathcal{R}_{\mu\nu\lambda}{}^\alpha 
\end{split}
\end{equation}
where  $\mathcal{R}_{\mu\nu\lambda}{}^\alpha$ is the curvature tensor associated with the Weyl-covariant derivative $\mathcal{D}_\lambda$ (See equation\eqref{covcomm:eq} in the next section).

\section{The Curvature tensors}\label{sec:curv}
To complete the classification of the various tensors that can be constructed at the second derivative level, we need to study the curvature tensors that appear via the commutators of two covariant derivatives. Hence, in this section, we consider in some detail the various curvature related observables in conformal hydrodynamics. In addition, we use this section to establish the notation for the various curvature tensors that appear in this paper.

We can define a curvature associated with the Weyl-covariant derivative by the usual procedure of evaluating the commutator between two covariant derivatives. The standard formalism goes through except for some subtleties we mention below. For a covariant vector field $V_\mu=\e^{-w\phi}\widetilde{V}_\mu\ $, we get
\begin{equation}\label{covcomm:eq}
\begin{split}
[\mathcal{D}_\mu,\mathcal{D}_\nu]V_\lambda &= w\ \mathcal{F}_{\mu\nu}\ V_\lambda - \mathcal{R}_{\mu\nu\lambda}{}^\alpha\  V_\alpha\quad \quad \text{with}\\ 
\mathcal{F}_{\mu\nu} &= \nabla_\mu \mathcal{A}_\nu - \nabla_\nu \mathcal{A}_\mu \\
\mathcal{R}_{\mu\nu\lambda}{}^\alpha &= R_{\mu\nu\lambda}{}^\alpha + \nabla_\mu \brk{{g}_{\lambda\nu}\mathcal{A}^{\alpha} - \delta^{\alpha}_{\lambda}\mathcal{A}_\nu  - \delta^{\alpha}_{\nu}\mathcal{A}_{\lambda}} - \nabla_\nu 
\brk{{g}_{\lambda\mu}\mathcal{A}^{\alpha} - \delta^{\alpha}_{\lambda}\mathcal{A}_\mu  - \delta^{\alpha}_{\mu}\mathcal{A}_{\lambda}}\\
&\quad +\brk{{g}_{\lambda\nu}\mathcal{A}^{\beta} - \delta^{\beta}_{\lambda}\mathcal{A}_\nu  - \delta^{\beta}_{\nu}\mathcal{A}_{\lambda}}
\brk{{g}_{\beta\mu}\mathcal{A}^{\alpha} - \delta^{\alpha}_{\beta}\mathcal{A}_\mu  - \delta^{\alpha}_{\mu}\mathcal{A}_{\beta}} \\
&\quad -\brk{{g}_{\lambda\mu}\mathcal{A}^{\beta} - \delta^{\beta}_{\lambda}\mathcal{A}_\mu  - \delta^{\beta}_{\mu}\mathcal{A}_{\lambda}}
\brk{{g}_{\beta\nu}\mathcal{A}^{\alpha} - \delta^{\alpha}_{\beta}\mathcal{A}_\nu  - \delta^{\alpha}_{\nu}\mathcal{A}_{\beta}}\\
\end{split}
\end{equation}
where we have introduced two new Weyl-invariant tensors $ \mathcal{F}_{\mu\nu}=\widetilde{\mathcal{F}}_{\mu\nu}$ and $\mathcal{R}_{\mu\nu\lambda}{}^\alpha=\widetilde{\mathcal{R}}_{\mu\nu\lambda}{}^\alpha$. The generalization to arbitrary tensors is straightforward.\footnote{As is evident from the notation above, we use calligraphic alphabets to denote the Weyl-covariant counterparts of the usual curvature tensors. Our notation for the usual Riemann tensor is defined by the relation 
\begin{equation}\label{gradcomm:eq}
[\nabla_\mu,\nabla_\nu]V^\lambda=R_{\mu\nu\sigma}{}^{\lambda}V^\sigma .
\end{equation}}

The above expression for $\mathcal{R}_{\mu\nu\lambda}{}^\alpha$ can be rewritten in the form
\begin{equation}\label{calcurv:eq}
\begin{split}
\mathcal{R}_{\mu\nu\lambda\sigma}&= R_{\mu\nu\lambda\sigma} + \delta^\alpha_{[\mu}g_{\nu][\lambda}\delta^\beta_{\sigma]}\left(\nabla_\alpha \mathcal{A}_\beta + \mathcal{A}_\alpha \mathcal{A}_\beta - \frac{\mathcal{A}^2}{2} g_{\alpha\beta} \right) - \mathcal{F}_{\mu\nu} g_{\lambda\sigma}
\end{split}
\end{equation}
where $B_{[\mu\nu]}\equiv B_{\mu\nu}- B_{\nu\mu}$ indicates antisymmetrisation. We can write down similar expressions involving Ricci tensor, Ricci scalar and Einstein tensor.
\begin{equation}\label{ricEin:eq}
\begin{split}
\mathcal{R}_{\mu\nu}&\equiv \mathcal{R}_{\mu\alpha\nu}{}^{\alpha} = R_{\mu\nu} -(d-2)\left(\nabla_\mu \mathcal{A}_\nu + \mathcal{A}_\mu \mathcal{A}_\nu -\mathcal{A}^2 g_{\mu\nu}  \right)-g_{\mu\nu}\nabla_\lambda\mathcal{A}^\lambda - \mathcal{F}_{\mu\nu} = \widetilde{\mathcal{R}}_{\mu\nu}\\
\mathcal{R} &\equiv \mathcal{R}_{\alpha}{}^{\alpha} = R -2(d-1)\nabla_\lambda\mathcal{A}^\lambda + (d-2)(d-1) \mathcal{A}^2 = \e^{-2\phi} \widetilde{\mathcal{R}}\\
\mathcal{G}_{\mu\nu} &\equiv \mathcal{R}_{\mu\nu}-\frac{\mathcal{R}}{2} g_{\mu\nu} = G_{\mu\nu} -(d-2)\brk{\nabla_\mu \mathcal{A}_\nu + \mathcal{A}_\mu \mathcal{A}_\nu - \left(\nabla_\lambda\mathcal{A}^\lambda-\frac{d-3}{2}\mathcal{A}^2\right)g_{\mu\nu}} - \mathcal{F}_{\mu\nu}\\
\end{split}
\end{equation}

These curvature tensors obey various Bianchi identities \footnote{These identities can be derived  from the Jacobi identity for the covariant derivative - $[\mathcal{D}_{[\mu},[\mathcal{D}_{\nu]},\mathcal{D}_\lambda]+[\mathcal{D}_\lambda,[\mathcal{D}_\mu,\mathcal{D}_\nu] ] = 0 $ }
\begin{equation}\label{bianchi:eq}
\begin{split}
\mathcal{R}_{\mu\nu\lambda}{}^\alpha + \mathcal{R}_{\lambda[\mu\nu]}{}^\alpha &=0 \\
\mathcal{D}_\lambda \mathcal{F}_{\mu\nu} + \mathcal{D}_{[\mu}\mathcal{F}_{\nu]\lambda} &=0 \\
\mathcal{D}_\lambda \mathcal{R}_{\mu\nu\alpha}{}^\beta + \mathcal{D}_{[\mu}\mathcal{R}_{\nu]\lambda\alpha}{}^\beta &=0
\end{split}
\end{equation}
and various reduced Bianchi identities\footnote{These identities are obtained from the Bianchi identities by contractions. }
\begin{equation}\label{redbianchi:eq}
\begin{split}
\mathcal{R}_{[\mu\nu]} =\mathcal{R}_{\mu\nu\alpha}{}^{\alpha} &= -d\ \mathcal{F}_{\mu\nu} \\
\mathcal{D}_{[\mu}\mathcal{R}_{\nu]\lambda} + \mathcal{D}_\sigma \mathcal{R}_{\mu\nu\lambda}{}^{\sigma} &= 0 \\
\mathcal{D}_{\lambda}\left(\mathcal{G}^{\mu\lambda}+\mathcal{F}^{\mu\lambda}\right) &=0
\end{split}
\end{equation}

The tensor $\mathcal{R}_{\mu\nu\lambda\sigma}$ does not have the same symmetry properties as that of the usual Riemann tensor. For example,
\begin{equation}\label{curvsym:eq}
\begin{split}
\mathcal{R}_{\mu\nu\lambda\sigma}+\mathcal{R}_{\mu\nu\sigma\lambda} &= -2\ \mathcal{F}_{\mu\nu} g_{\lambda\sigma} \\
\mathcal{R}_{\mu\nu\lambda\sigma}-\mathcal{R}_{\lambda\sigma\mu\nu} &= \delta^\alpha_{[\mu}g_{\nu][\lambda}\delta^\beta_{\sigma]} \mathcal{F}_{\alpha\beta} - \mathcal{F}_{\mu\nu} g_{\lambda\sigma} + \mathcal{F}_{\lambda\sigma} g_{\mu\nu} \\
\mathcal{R}_{\mu\alpha\nu\beta}V^\alpha V^\beta-\mathcal{R}_{\nu\alpha\mu\beta}V^\alpha V^\beta &= - \mathcal{F}_{\mu\nu}\ V^\alpha V_\alpha\\
\end{split}
\end{equation}

The conformal tensors of the underlying spacetime manifold appear in the above formalism as a subset of conformal observeables in hydrodynamics. These conformal tensors are the Weyl-covariant tensors that are independent of the background fluid velocity. The Weyl curvature $C_{\mu\nu\lambda\sigma}$ is a well-known example of a conformal tensor. We have(for $d\geq3$)
\begin{equation}\label{weylcurv:eq}
\begin{split}
\mathcal{C}_{\mu\nu\lambda\sigma} &\equiv \mathcal{R}_{\mu\nu\lambda\sigma}+\delta^\alpha_{[\mu}g_{\nu][\lambda}\delta^\beta_{\sigma]}\mathcal{S}_{\alpha\beta} = C_{\mu\nu\lambda\sigma} - \mathcal{F}_{\mu\nu} g_{\lambda\sigma} = \e^{2\phi}\widetilde{\mathcal{C}}_{\mu\nu\lambda\sigma}
\end{split}
\end{equation}
where the Schouten tensor $\mathcal{S}_{\mu\nu}$ is defined as\footnotemark
\begin{equation}\label{schouten:eq}
\begin{split}
\mathcal{S}_{\mu\nu} &\equiv \frac{1}{d-2}\left(\mathcal{R}_{\mu\nu}-\frac{\mathcal{R}g_{\mu\nu}}{2(d-1)}\right) = S_{\mu\nu}-\left(\nabla_\mu \mathcal{A}_\nu + \mathcal{A}_\mu \mathcal{A}_\nu - \frac{\mathcal{A}^2}{2} g_{\mu\nu} \right) -\frac{\mathcal{F}_{\mu\nu}}{d-2} =\widetilde{\mathcal{S}}_{\mu\nu}
\end{split}
\end{equation}

\footnotetext{Often in the study of conformal tensors , it is useful to rewrite other curvature tensors in terms of the Schouten and the Weyl curvature tensors-
\begin{equation}\label{weylschouten:eq}
\begin{split}
\mathcal{R}_{\mu\nu\lambda\sigma}&=\mathcal{C}_{\mu\nu\lambda\sigma} -\delta^\alpha_{[\mu}g_{\nu][\lambda}\delta^\beta_{\sigma]}\mathcal{S}_{\alpha\beta},\qquad \mathcal{R}= 2(d-1)\mathcal{S}_\lambda{}^\lambda \\
\mathcal{R}_{\mu\nu}&= (d-2) \mathcal{S}_{\mu\nu} + \mathcal{S}_\lambda{}^\lambda g_{\mu\nu} ,\qquad  \mathcal{G}_{\mu\nu}= (d-2)(\mathcal{S}_{\mu\nu}-\mathcal{S}_\lambda{}^\lambda g_{\mu\nu}) \\
\end{split}
\end{equation}}

From equation \eqref{weylcurv:eq}, it is clear that $ C_{\mu\nu\lambda\sigma}=\mathcal{C}_{\mu\nu\lambda\sigma} +\mathcal{F}_{\mu\nu} g_{\lambda\sigma} $ is clearly a conformal tensor. Such an analysis can in principle be repeated for the other known conformal tensors in arbitrary dimensions.


The Weyl Tensor $ C_{\mu\nu\lambda\sigma}$ has the same symmetry properties as that of Riemann Tensor $ R_{\mu\nu\lambda\sigma}$.
\begin{equation}\label{weylsym:eq}
\begin{split}
C_{\mu\nu\lambda\sigma} &= -C_{\nu\mu\lambda\sigma} = -C_{\mu\nu\sigma\lambda}=C_{\lambda\sigma\mu\nu} \\
\qquad &\text{and}\quad C_{\mu\alpha\lambda}{}^\alpha =0
\end{split}
\end{equation}
From which it follows that $C_{\mu\alpha\nu\beta}u^\alpha u^\beta$ is a symmetric traceless and transverse tensor - a fact which will turn out to be important later in our discussion of conformal hydrodynamics.

\section{Conformal hydrodynamics}\label{sec:confhydro}
In this section, we reformulate the fundamental equations of fluid mechanics in a Weyl-covariant form. The basic equations of fluid mechanics are the conservation of energy-momentum and various other charges -
\begin{equation}\label{flumech:eq}
\begin{split}
\nabla_\mu T^{\mu\nu} = 0  \qquad \text{and} \qquad \nabla_\mu J^\mu=0 
\end{split}
\end{equation}

But, these equations are not manifestly Weyl-covariant. To cast them into a manifestly Weyl-covariant form, we need the transformation of the stress tensor and the currents - $T^{\mu\nu} = \e^{-(d+2)\phi} \widetilde{T}^{\mu\nu} + \ldots $ and $J^\mu = \e^{-w \phi} \tilde{J}^\mu$ respectively (where $\ldots$ denotes the contributions due to the Weyl anomaly $T^\lambda{}_\lambda = \mathcal{W}$. The Weyl Anomaly $\mathcal{W}$ only on the microscopic field content and the ambient spacetime in which the conformal fluid lives.). Then, we can impose a manifestly Weyl covariant\footnote{The Weyl transformation of the stress tensor in quantum theories is non-trivial because of the presence of Weyl anomaly . The situation is simplified if we assume that there exists a symmetric tensor $T^{\mu\nu}_{\text{conf}}= T^{\mu\nu} -\mathcal{W}^{\mu\nu}[g]= \e^{-(d+2)\phi} \widetilde{T}^{\mu\nu}_{\text{conf}}$ where $\mathcal{W}^{\mu\nu}[g]$ characterizes the contribution due to Weyl anomaly which depends only on the background spacetime and the field content. In that case, though $T^{\mu\nu}$ does not transform homogeneously under the Weyl transformations, one can show that $\mathcal{D}_\mu T^{\mu\nu}=\e^{-(d+2)\phi}\widetilde{\mathcal{D}}_\mu \tilde{T}^{\mu\nu}$ with $\mathcal{D}_\mu T^{\mu\nu}$ defined as above. This shows that the contributions due to Weyl anomaly can be taken into account with slight modifications. In what follows, we will ignore such subtleties due to Weyl anomaly - we will just assume that the energy-momentum tensor is traceless with the presumption that the statements we make can always be suitably modified once trace anomaly is taken into account. } set of equations
\begin{equation}
\begin{split}\label{confflu:eq}
\mathcal{D}_\mu T^{\mu\nu} &= \nabla_\mu T^{\mu\nu} + \mathcal{A}^\nu (T^\mu{}_\mu-\mathcal{W}) =0\\
\mathcal{D}_\mu J^\mu &= \nabla_\mu J^\mu +(w-d) \mathcal{A}_\mu J^\mu=0
\end{split}
\end{equation}
These equations coincide with \eqref{flumech:eq} provided $T^{\mu\nu}$ is a traceless tensor of conformal weight $d+2$ apart from the anomalous contribution and the conformal weight $w$ of the conserved current is equal to the number of dimensions of the spacetime. The second condition is same as requiring that the charge associated with the charge currents be a dimensionless scalar. 

The entropy current $J^\mu_S$ of the fluid also has a conformal weight equal to the spacetime dimensions. This means that we can write the statement of the second law in a manifestly conformal way as  
\begin{equation}\label{secondlaw:eq}
\begin{split}
\mathcal{D}_\mu J^\mu_S = \nabla_\mu J^\mu_S \geq 0
\end{split}
\end{equation}
Similarly, the first law of thermodynamics $\tloc u^\lambda\nabla_\lambda s= (d-1) u^\lambda\nabla_\lambda p  - \mu_i u^\lambda\nabla_\lambda \rho_i$ can be written in a conformal form 
\begin{equation}\label{firstlaw:eq}
\tloc u^\lambda\mathcal{D}_\lambda s= (d-1) u^\lambda\mathcal{D}_\lambda p  - \mu_i u^\lambda\mathcal{D}_\lambda \rho_i
\end{equation}
where $(d-1)p$ is the energy density of the conformal fluid. \footnote{Note that the additional terms  that appear when one converts $\nabla$ to $\mathcal{D}$ in \eqref{firstlaw:eq} cancel out because of Gibbs-Duhem Relation $\tloc s= (d-1)p + p  - \mu_i \rho_i$ where $(d-1)p$ is the energy density of the conformal fluid.}

The fluid mechanics is completely specified once the expressions of the energy momentum tensor, the charged currents and the entropy current in terms of the velocity, temperature and the chemical potentials. The conventional discussion on relativistic hydrodynamics(say as given by Landau and Lifshitz\cite{1959flme.book.....L}) can be adopted to the case of conformal fluids with the additional condition that the energy momentum tensor of a conformal fluid is traceless. The energy-momentum tensor, the charged currents and the entropy current of the fluid are usually divided into a non-dissipative part and a dissipative part.  
\begin{equation}\label{dissdef:eq}
\begin{split}
T^{\mu\nu} &= \ p \left( g^{\mu \nu} + d\ u^\mu u^\nu \right) + \pi^{\mu\nu}\\
J^\mu_i &= \rho_i u^\mu + \nu^\mu_i \\
J^\mu_S &= s u^\mu + J^\mu_{S,\text{diss}}
\end{split}
\end{equation}
where we take the visco-elastic stress $\pi^{\mu\nu}$ to be transverse   $(u_\mu \pi^{\mu\nu} = 0)$ and traceless $(\pi^\mu{}_\mu=0)$ and the diffusion current $\nu^\mu_i$ to be transverse $(u_\lambda \nu_i^{\lambda} = 0)$. This in turn implies the following equations
\begin{equation}\label{doteq:eq}
\begin{split}
0 &= -u_\nu \mathcal{D}_\mu T^{\mu\nu} = (d-1) u^\lambda\mathcal{D}_\lambda p + \pi^{\mu\nu}\sigma_{\mu\nu} \\
0 & = \mathcal{D}_\lambda J^{\lambda}_i = u^\lambda\mathcal{D}_\lambda \rho_i +\mathcal{D}_\lambda \nu_i^{\lambda}
\end{split}
\end{equation}

We can now use the first law of thermodynamics \eqref{firstlaw:eq} to conclude
\begin{equation}\label{sdiss:eq}
\begin{split}
\tloc \mathcal{D}_\mu J^\mu_S = -  \pi^{\mu\nu}\sigma_{\mu\nu} + \mu_i\mathcal{D}_\lambda \nu_i^{\lambda} +  \tloc {\mathcal{D}}_\mu J^\mu_{S,\text{diss}} \geq 0
\end{split}
\end{equation}

Now we can write down the most general form of the dissipative currents confining ourselves to no more than second derivatives in velocity.\footnote{Given the fact that for a conformal fluid $p\sim \tloc^d$ and the equation of motion $u^\lambda\mathcal{D}_\lambda p \sim \pi^{\mu\nu}\sigma_{\mu\nu}$ we conclude that wherever a single derivative of $\tloc$ occurs, it can be replaced by a term involving two or more derivatives of the fluid velocity. Hence, for the sake of counting, one derivative of $\tloc$ should be counted as equivalent to two derivatives of $u^\mu$. } For simplicity, we will consider here the case when no charges are present - the generalization to the case when there are conserved charges is straightforward. Hence, a general derivative expansion for the energy-momentum tensor $T^{\mu\nu}$ is given by
\begin{equation}\label{derivexp:eq}
\begin{split}
T^{\mu\nu}&= \eta_0 \tloc^{d} (g^{\mu\nu}+d u^\mu u^\nu) \\
&\quad +\eta_1 \tloc^{d-1} \sigma^{\mu\nu}  \\
&\quad + \eta_2 \tloc^{d-2} \ u^\lambda \mathcal{D}_\lambda \sigma^{\mu\nu}+\eta_3\ \tloc^{d-2}  [\omega^\mu{}_\lambda\sigma^{\lambda\nu}+\omega^\nu{}_\lambda\sigma^{\lambda\mu}] \\
&\quad +\eta_4\ \tloc^{d-2} [\sigma^\mu{}_\lambda\sigma^{\lambda\nu}-\frac{P^{\mu\nu}}{d-1}\sigma^{\alpha\beta}\sigma_{\alpha\beta}]+\eta_5\ \tloc^{d-2}  [\omega^\mu{}_\lambda\omega^{\lambda\nu}+\frac{P^{\mu\nu}}{d-1}\omega^{\alpha\beta}\omega_{\alpha\beta}]\\
&\quad  +\eta_6\ \tloc^{d-2} C^{\mu}{}_{\alpha}{}^{\nu}{}_{\beta}u^\alpha u^\beta
\end{split}
\end{equation}
where the first line denotes the non-dissipative part(with the conformal equation of state $p=\eta_0 \tloc^d$) and the rest denote the visco-elastic stress $\pi^{\mu\nu}$. We show in the appendix (\ref{sec:cfnofR}) that no more terms appear at this order in the derivative expansion. This derivative expansion in terms of conformally covariant terms was first analyzed in \cite{Baier:2007ix} and our discussion here closely parallels theirs.\footnote{Refer \S \ref{sec:adsfluid} to see how our notation is related to that of \cite{Bhattacharyya:2007jc} and \cite{Baier:2007ix}.}.  

\section{Entropy current in Conformal hydrodynamics}\label{sec:entropy}
Now we can write down the expression for the second law by restricting \eqref{sdiss:eq} to the case where there are no charges, and then substituting for $\pi^{\mu\nu}$ from \eqref{derivexp:eq}
\begin{equation}\label{sdissderiv:eq}
\begin{split}
\tloc \mathcal{D}_\mu J^\mu_S &=  \tloc {\mathcal{D}}_\mu J^\mu_{S,\text{diss}} -\eta_1 \tloc^{d-1} \sigma^{\mu\nu} \sigma_{\mu\nu} - \eta_2 \tloc^{d-2}\sigma_{\mu\nu} \ u^\lambda \mathcal{D}_\lambda \sigma^{\mu\nu} \\
&\quad -\eta_4\ \tloc^{d-2}\sigma_{\mu\nu} \sigma^\mu{}_\lambda\sigma^{\lambda\nu}-\eta_5\ \tloc^{d-2} \sigma_{\mu\nu}\omega^\mu{}_\lambda\omega^{\lambda\nu}\\
&\quad  -\eta_6\ \tloc^{d-2}\sigma^{\mu\nu}C_{\mu\alpha\nu\beta}u^\alpha u^\beta
\end{split}
\end{equation}

Now we invoke two identities(see appendix \ref{sec:identity} for the proofs)
\begin{equation}\label{theidentity:eq}
\begin{split}
\sigma^{\mu\nu} \omega_\mu{}^\alpha \omega_{\alpha\nu} &= \mathcal{D}_\lambda\brk{\frac{\omega^{\mu\nu}\omega_{\mu\nu}}{4}u^\lambda+
\frac{\mathcal{D}_\nu \omega^{\lambda\nu}}{2(d-3)}}\\
\sigma^{\mu\nu} C_{\mu\alpha\nu\beta} u^\alpha u^\beta  &= \sigma^{\mu\nu} \sigma_\mu{}^\alpha \sigma_{\alpha\nu} + \mathcal{D}_\lambda\brk{ \frac{2\sigma^{\mu\nu}\sigma_{\mu\nu}+\omega^{\mu\nu}\omega_{\mu\nu}}{4} u^\lambda+\frac{u_\mu(\mathcal{G}^{\mu\lambda}+\mathcal{F}^{\mu\lambda})}{d-2}+
\frac{3\mathcal{D}_\nu \omega^{\lambda\nu}}{2(d-3)}}\\
\end{split}
\end{equation}
to write
\begin{equation}\label{sdissderiv2:eq}
\begin{split}
\tloc \mathcal{D}_\mu J^\mu_S &=  -\eta_1 \tloc^{d-1} \sigma^{\mu\nu} \sigma_{\mu\nu}-(\eta_4+\eta_6)\ \tloc^{d-2}\sigma_{\mu\nu} \sigma^\mu{}_\lambda\sigma^{\lambda\nu} + \tloc {\mathcal{D}}_\mu J^\mu_{S,\text{diss}} \\
&  - \tloc^{d-2}\mathcal{D}_\lambda\left[\left(\frac{2(\eta_2+\eta_6)\ \sigma^{\mu\nu} \sigma_{\mu\nu}+(\eta_5+\eta_6)\ \omega^{\mu\nu} \omega_{\mu\nu}}{4}\right) u^\lambda\right.\\
&\quad\left. + \frac{\eta_6\ u_\mu(\mathcal{G}^{\mu\lambda}+\mathcal{F}^{\mu\lambda})}{d-2}+\frac{(\eta_5+3\eta_6)}{2(d-3)}\mathcal{D}_\nu\omega^{\lambda\nu}\right]\\
\end{split}
\end{equation}

We now want to propose an expression for the dissipative entropy flux such that the total entropy obeys the second law of thermodynamics. In this paper, we give a specific proposal for this entropy current which is consistent with the second law.\footnote{Note that, the second law alone does not determine the entropy flux uniquely - for example, an additional term with positive divergence can always be added to the dissipative entropy flux without violating the second law. Given this fact, it is important to emphasize that what is being proposed here is just one possible definition of the entropy current. See \S\ref{sec:disc} for a discussion of this issue.}  Taking the dissipative entropy flux as
\begin{equation}\label{jsdiss:eq}
\begin{split}
J^\lambda_{S,\text{diss}} &= \left(\frac{2(\eta_2+\eta_6)\tloc^{d-3}\ \sigma^{\mu\nu} \sigma_{\mu\nu}+(\eta_5+\eta_6)\tloc^{d-3}\ \omega^{\mu\nu} \omega_{\mu\nu}}{4}\right) u^\lambda\\
&\quad + \frac{\eta_6\tloc^{d-3}\ u_\mu(\mathcal{G}^{\mu\lambda}+\mathcal{F}^{\mu\lambda})}{d-2}+\frac{(\eta_5+3\eta_6)\tloc^{d-3}}{2(d-3)}\mathcal{D}_\nu\omega^{\lambda\nu}\\
\end{split}
\end{equation}
and keeping only terms with three derivatives or less of velocity\footnote{Since we are working with the divergence of quantities accurate up to second derivatives of velocity, consistency demands that we keep terms involving three derivatives or less. Further, as before, we use the equations of motion to replace a derivative of $\tloc$ by a term involving two or more derivatives of the fluid velocity.}
\begin{equation}\label{sdissfinal:eq}
\begin{split}
\tloc \mathcal{D}_\mu J^\mu_S &=  -\eta_1 \tloc^{d-1} \sigma^{\mu\nu} \sigma_{\mu\nu}  -(\eta_4+\eta_6)\ \tloc^{d-2}\sigma_{\mu\nu} \sigma^\mu{}_\lambda\sigma^{\lambda\nu}\\
&= -\eta_1 \tloc^{d-1} \brk{\sigma^{\mu\nu}+\frac{\eta_4+\eta_6}{2\eta_1\tloc} \sigma^\mu{}_\lambda\sigma^{\lambda\nu}}\brk{\sigma_{\mu\nu}+\frac{\eta_4+\eta_6}{2\eta_1\tloc} \sigma_\mu{}^\alpha\sigma_{\alpha\nu}}
\end{split}
\end{equation}
from which we conclude that 
\begin{equation}\label{etacondn:eq}
\begin{split}
\eta_1 \leq 0\ 
\end{split}
\end{equation}
along with a dissipative current of the form given in equation\eqref{jsdiss:eq} is sufficient to ensure that the conformal fluid obeys the second law\footnote{This section has greatly benefited from my discussions with Shiraz Minwalla regarding the validity of second law for the entropy flux proposed above. I would also like to thank  Veronica Hubeny, Giuseppe Policastro, Mukund Rangamani, Dam Thonh Son and Misha Stephanov for commenting on an earlier version of this section.} 
\begin{equation}\label{secondconf:eq}
\begin{split}
\tloc \mathcal{D}_\mu J^\mu_S &=   -\eta_1 \tloc^{d-1} \brk{\sigma^{\mu\nu}+\frac{\eta_4+\eta_6}{2\eta_1\tloc} \sigma^\mu{}_\lambda\sigma^{\lambda\nu}}\brk{\sigma_{\mu\nu}+\frac{\eta_4+\eta_6}{2\eta_1\tloc} \sigma_\mu{}^\alpha\sigma_{\alpha\nu}}
 \geq 0
\end{split}
\end{equation}

Hence for a general energy-momentum tensor of the form
\begin{equation}\label{Tfinal:eq}
\begin{split}
T^{\mu\nu}&= p (g^{\mu\nu}+d u^\mu u^\nu) \\
&\quad -2\eta \brk{\sigma^{\mu\nu}-\tau_{\pi}\ u^\lambda \mathcal{D}_\lambda \sigma^{\mu\nu}+\tau_\omega(\omega^\mu{}_\lambda\sigma^{\lambda\nu}+\omega^\nu{}_\lambda\sigma^{\lambda\mu}) }\\
&\quad +\xi_\sigma [\sigma^\mu{}_\lambda\sigma^{\lambda\nu}-\frac{P^{\mu\nu}}{d-1}\sigma^{\alpha\beta}\sigma_{\alpha\beta}]-\xi_C\ C_{\mu\alpha\nu\beta}u^\alpha u^\beta\\
&\quad +\xi_\omega [\omega^\mu{}_\lambda\omega^{\lambda\nu}+\frac{P^{\mu\nu}}{d-1}\omega^{\alpha\beta}\omega_{\alpha\beta}]\\
\end{split}
\end{equation}
where we have defined 
\begin{equation}\label{etaidef:eq}
\begin{split}
p &= \eta_0\tloc^d, \quad -2\eta=\eta_1\tloc^{d-1}, \quad 2\eta\tau_{\pi}=\eta_2\tloc^{d-2} \\ 
-2\eta\tau_{\omega}&=\eta_3\tloc^{d-2},\quad \xi_\sigma=\eta_4\tloc^{d-2},\quad \xi_C=-\eta_6\tloc^{d-2}, \quad \xi_\omega=\eta_5\tloc^{d-2}\\
\end{split}
\end{equation}
the proposed expression for the entropy current is 
\begin{equation}\label{jsfinal:eq}
\begin{split}
J^\lambda_s &= s u^\lambda + J^\lambda_{S,\text{diss}} \\
&= \left( s-\frac{2(\xi_C-2\eta\tau_{\pi})\ \sigma^{\mu\nu} \sigma_{\mu\nu}+(\xi_C-\xi_\omega)\ \omega^{\mu\nu} \omega_{\mu\nu}}{4\tloc}\right) u^\lambda\\
&\quad - \frac{\xi_C u_\mu(\mathcal{G}^{\mu\lambda}+\mathcal{F}^{\mu\lambda})}{(d-2)\tloc}-\frac{(3\xi_C-\xi_\omega)}{2(d-3)\tloc}\mathcal{D}_\nu\omega^{\lambda\nu}\\
\qquad \text{with}\qquad &\tloc \mathcal{D}_\mu J^\mu_S = 2\eta \brk{\sigma^{\mu\nu}+\frac{\xi_C-\xi_\sigma}{4\eta} \sigma^\mu{}_\lambda\sigma^{\lambda\nu}}\brk{\sigma_{\mu\nu}+\frac{\xi_C-\xi_\sigma}{4\eta}\sigma_\mu{}^\alpha\sigma_{\alpha\nu}} \geq 0
\end{split}
\end{equation}

These expressions completely determine the dynamics of a conformal fluid up to second derivatives in the derivative expansion.  We now proceed to apply the above formalism to the constitutive relations of $\mathcal{N}=4$ SYM fluid derived recently using AdS/CFT correspondence.  

\section{$\mathcal{N}=4$ SYM fluid : Energy-momentum and Entropy current }\label{sec:adsfluid}
A prominent example of a conformal fluid in four dimensions is the fluid made out of the matter content in $\mathcal{N}=4$ supersymmetric Yang-Mills theory. The flat spacetime stress tensor for the four dimensional conformal fluids with AdS duals (which in particular includes $\mathcal{N}=4$ SYM fluid in the four dimensional Minkowski spacetime) has been calculated recently via AdS/CFT upto second derivative terms \cite{Bhattacharyya:2007jc}. Independently, in \cite{Baier:2007ix}, its authors wrote down the general derivative expansion for a conformal fluid and determined some of the coefficients occurring in that expansion. In this section, we relate the work done in above references to the formalism developed here.

The expression for the energy-momentum tensor derived in \cite{Bhattacharyya:2007jc} is
\begin{equation} \label{st:eq}
\begin{split}
T^{\mu\nu}&=\ p \left( g^{\mu \nu} + 4 u^\mu u^\nu \right)\\
&\quad - 2\ \eta\ \sigma^{\mu\nu}  + 2\ \eta\  \frac{\left(\ln 2\right) T_{2a}^{\mu \nu} +2\ T_{2b}^{\mu\nu} +  \left( 2- \ln 2 \right)  \left[ \frac{1}{3} T_{2c}^{\mu \nu}
+ T_{2d}^{\mu\nu} + T_{2e}^{\mu\nu} \right] }{2\pi\tloc} \\
\end{split}
\end{equation}
where
\begin{equation}\label{stdef:eq}
\begin{split}
p &= \frac{N_c^2}{8\pi^2}(\pi\tloc)^4  ;\qquad \eta = \frac{N_c^2}{8\pi^2}(\pi\tloc)^3\\
\vartheta &= \nabla_\lambda  u^\lambda\ ;\qquad  a^\mu  =  u^\lambda \nabla_\lambda  u^\mu ; \quad {l}_\mu =\epsilon_{\alpha \beta \gamma \mu} u^\alpha \omega^\beta{}^\gamma ;\\
\sigma^{\mu\nu}&= P^{\mu \alpha} P^{\nu \beta}
\left( \frac{ \nabla_\alpha u_\beta + \nabla_\beta u_\alpha}{2}\right)
-P^{\mu \nu} \frac{\nabla_\alpha u^\alpha}{3} ;\\
T_{2a}^{\mu\nu}&= \frac{\epsilon^{\alpha \beta \gamma \mu}
u_\alpha {l}_\beta \sigma_\gamma{}^\nu  + \epsilon^{\alpha \beta \gamma\nu}
u_\alpha {l}_\beta \sigma_\gamma{}^\mu }{2} ;\\
T_{2b}^{\mu\nu}&= \sigma^{\mu\alpha} \sigma_{\alpha}^{\nu} - \frac{P^{\mu\nu}}{3}\sigma^{\beta \alpha} \sigma_{\alpha \beta} ; \\
T_{2c}^{\mu\nu}&=\vartheta \sigma^{\mu\nu} ;\quad T_{2d}^{\mu\nu} = a^\mu  a^\nu  -  a_\lambda a^\lambda \frac{P^{\mu\nu}}{3} ;\\
T_{2e}^{\mu \nu}&= P^{\mu\alpha} P^{\nu\beta} u^\lambda \nabla_\lambda \left(
  \frac{ \nabla_\alpha u_\beta + \nabla_\beta u_\alpha}{2} \right)
- \frac{P^{\mu \nu}} {3} P^{\beta \gamma} u^\lambda \nabla_\lambda \left(\nabla_\beta u_\gamma \right)  ;
\end{split}
 \end{equation}
where $\epsilon_{0123} =-\epsilon^{0123} =  1$ and we are working in flat co-ordinates of the Minkowski spacetime. The above expression can be rewritten in terms of manifestly conformal observables as follows.
\begin{equation}\label{stredef:eq}
\begin{split}
T_{2a}^{\mu\nu}&= - \omega^{\mu}{}_\lambda\sigma^{\lambda\nu}-\omega^{\nu}{}_\lambda\sigma^{\lambda\mu} \qquad , \qquad 
T_{2b}^{\mu\nu}= \sigma^{\mu\alpha} \sigma_{\alpha}{}^{\nu} - \frac{P^{\mu\nu}}{3}\sigma^{\beta \alpha} \sigma_{\alpha \beta}  \\
\frac{1}{3} T_{2c}^{\mu \nu}+ T_{2d}^{\mu\nu} + T_{2e}^{\mu\nu} &= P^{\mu\alpha}P^{\nu\beta}u^{\lambda}\nabla_{\lambda}{\sigma}_{\alpha\beta} + \frac{\vartheta}{d-1}{\sigma}_{\mu\nu}=P^{\mu\alpha}P^{\nu\beta}u^{\lambda}\mathcal{D}_{\lambda} {\sigma}_{\alpha\beta}=u^{\lambda}\mathcal{D}_{\lambda} \sigma^{\mu\nu}
\end{split}
\end{equation}
The stress tensor becomes
\begin{equation} \label{stre:eq}
\begin{split}
T^{\mu\nu} &=\ p \left( g^{\mu \nu} + 4 u^\mu u^\nu \right)  \\
&\quad - 2\ \eta\ \left[\sigma^{\mu \nu} - \frac{\left( 2- \ln 2 \right)}{2\pi \tloc} u^{\lambda}\mathcal{D}_{\lambda} \sigma^{\mu\nu}+ \frac{\left(\ln 2\right)}{2\pi\tloc} (\omega^{\mu}{}_\lambda\sigma^{\lambda\nu}+\omega^{\nu}{}_\lambda\sigma^{\lambda\mu})\right] \\
&\qquad +  \frac{4\ \eta}{2\pi\tloc}[\sigma^{\mu\lambda} \sigma_{\lambda}{}^{\nu} - \frac{P^{\mu\nu}}{3}\sigma^{\alpha\beta} \sigma_{\alpha \beta}] \\
\end{split}
\end{equation}

This expression matches\footnote{Note that the calculation in \cite{Bhattacharyya:2007jc} was done for flat spacetime and hence the curvature term does not appear in their derivation.} with the expression in \eqref{Tfinal:eq} provided we take
\begin{equation}\label{symparam:eq}
\begin{split}
p = \frac{N_c^2}{8\pi^2}(\pi\tloc)^4\  &;\qquad \eta = \frac{N_c^2}{8\pi^2}(\pi\tloc)^3\ ;\\
\qquad \tau_\pi = \frac{ 2- \ln 2 }{2\pi \tloc}\ ;\qquad \tau_\omega = \frac{\ln 2}{2\pi\tloc}\ &;\qquad \xi_\sigma = \xi_C=\frac{4\ \eta}{2\pi\tloc}\ ;\qquad \xi_\omega = 0\ .
\end{split}
\end{equation}
where we have also included the value of the curvature coupling $\xi_C$ which was calculated by the authors of \cite{Baier:2007ix}.

Now, we proceed to compare the results of \cite{Baier:2007ix} to the results derived here. Translated into notations of this paper\footnote{Note that the $\sigma_{\mu\nu}$ of \cite{Baier:2007ix} is twice that of ours and their curvature tensors are negative of the curvature tensors defined in this paper.}  their expression (See Eqn.(3.11) of \cite{Baier:2007ix}) reads
\begin{equation}\label{strbaier:eq}
\begin{split}
\pi^{\mu\nu} =& -2\eta \sigma^{\mu\nu}+2\eta\tau_{\pi}\ u^\lambda \mathcal{D}_\lambda \sigma^{\mu\nu} -\kappa [P^{\mu\lambda}P^{\nu\sigma}R_{\lambda\sigma}+ (d-2) P^{\mu\lambda} P^{\nu\sigma} R_{\lambda\alpha\sigma\beta}u^{\alpha}u^{\beta} \\
&\quad - \frac{P^{\mu\nu}}{d-1}(P^{\lambda\sigma}R_{\lambda\sigma}+ (d-2) P^{\lambda\sigma} R_{\lambda\alpha\sigma\beta}u^{\alpha}u^{\beta})]\\
&\quad +4\lambda_1 (\sigma^\mu{}_\lambda \sigma^{\lambda\nu} - \frac{P^{\mu\nu}}{d-1} \sigma^{\alpha\beta} \sigma_{\alpha\beta}) + 4 \lambda_2 (\omega^\mu{}_\lambda\sigma^{\lambda\nu}+\omega^\nu{}_\lambda\sigma^{\lambda\mu}) \\
&\quad +\lambda_3 (\omega^\mu{}_\lambda \omega^{\lambda\nu} + \frac{P^{\mu\nu}}{d-1} \omega^{\alpha\beta}\omega_{\alpha\beta})\\
\end{split}
\end{equation}
with 
\begin{equation*}
p = \frac{N_c^2}{8\pi^2}(\pi\tloc)^4\  ;\qquad \eta = \frac{N_c^2}{8\pi^2}(\pi\tloc)^3\ ;\\
\qquad \tau_\pi = \frac{ 2- \ln 2 }{2\pi \tloc}\ ;\qquad \lambda_1=
\frac{\eta}{2\pi\tloc}\ ;\qquad \kappa=\frac{\eta}{\pi\tloc}\ ;\qquad  
\end{equation*}
and the parameters $\lambda_{2,3}$ were left undetermined in \cite{Baier:2007ix}. By inspection, we conclude that the above expression satisfies\footnote{ We have invoked the identity (which follows by applying projection operators to the the definition of Weyl tensor in \eqref{weylcurv:eq})
\begin{equation*}
\begin{split}
P^{\mu\lambda}P^{\nu\sigma}R_{\lambda\sigma}&+ (d-2) P^{\mu\lambda} P^{\nu\sigma} R_{\lambda\alpha\sigma\beta}u^{\alpha}u^{\beta} - \frac{P^{\mu\nu}}{d-1}(P^{\lambda\sigma}R_{\lambda\sigma}+ (d-2) P^{\lambda\sigma} R_{\lambda\alpha\sigma\beta}u^{\alpha}u^{\beta}) \\
&= (d-2) C_{\mu\alpha\nu\beta}u^\alpha u^\beta
\end{split}
\end{equation*}}
the conditions we laid down in \eqref{etacondn:eq}.The above expression is completely consistent with the coefficients we  derived above in \eqref{symparam:eq}. Hence, the second-order hydrodynamics of $\mathcal{N}=4$ SYM fluid is completely summarized by \eqref{symparam:eq}.

Now, we can use the discussion in our previous section to calculate the entropy current for $\mathcal{N}=4$ SYM fluid. Using the equation of state $\tloc s=p\ d=4p=4\pi\eta\tloc$ for a conformal fluid and \eqref{jsfinal:eq} we get 
\begin{equation}
\begin{split}\label{SYMJs:eq}
J^\lambda_s &=4\pi\eta \left[u^\lambda -\frac{\left[(\ln {2}) \sigma^{\mu\nu} \sigma_{\mu\nu}+\omega^{\mu\nu} \omega_{\mu\nu}\right]u^\lambda+2 u_\mu(\mathcal{G}^{\mu\lambda}+\mathcal{F}^{\mu\lambda})+6\mathcal{D}_\nu\omega^{\lambda\nu}}{8(\pi\tloc)^2}\right].\\
\qquad \text{with}\qquad &\tloc \mathcal{D}_\mu J^\mu_S =  2\eta \sigma^{\mu\nu} \sigma_{\mu\nu} \geq 0\\
\end{split}
\end{equation}
This expression gives the the next to leading order corrections to the holographic result $ J^\lambda_s =4\pi\eta u^\lambda $ of Kovtun, Son and Starinets\cite{Kovtun:2004de}. 

Note that our proposal for the entropy current was motivated in an indirect way - by first finding the holographic energy-momentum tensor and then guessing the entropy current from it by demanding second law. It would be interesting to do a direct gravity computation of the entropy current that checks this proposal. See \S \ref{sec:disc} for a discussion on this issue. Further, the rate of entropy production takes a very simple form in the case of $\mathcal{N}=4$ SYM fluid - the total entropy production is completely given by a term quadratic in shear strain rate $\sigma_{\mu\nu}$ and there is no contribution at the next order. This fact can be traced to an interesting fact that $\xi_\sigma=\xi_C$ for $\mathcal{N}=4$ SYM. 

We would now like to give a heuristic reason for why we might expect the entropy production to take such a simpler form. Notice that the additional contribution to the entropy production(over and above the standard shear viscosity part) comes from a viscoelastic stress of the form $\pi^{\mu\nu}\sim\sigma^{\mu}{}_\lambda \sigma^{\lambda\nu}$. The rate of energy transfer by such a stress is $\sigma_{\mu\nu}\pi^{\mu\nu}\sim \sigma_{\mu\nu}\sigma^{\mu}{}_\lambda \sigma^{\lambda\nu}$ . If this energy transfer was irreversible, this would contribute to an entropy production $-\tloc^{-1} \sigma_{\mu\nu}\pi^{\mu\nu}$ which is precisely the term which we arrived at in the last section.

However, the energy transfer by a stress of the form $\pi\sim \sigma\sigma$ is reversible - in particular, for such a stress, the rate of work done $\pi\sigma$ reverses sign if we reverse the fluid flow. If we assume that such a reversible energy transfer cannot contribute to entropy production, then either such a term can be absorbed into a redefinition of the $J^\mu_{\text{S,diss}}$ or the coefficient of such a contribution should vanish. The second possibility immediately yields the condition $\xi_\sigma=\xi_C$. This, however is a very heuristic line of reasoning and it would be interesting to know how far it is valid. In principle, it should be possible to extend the holographic calculation of $\xi_C$ and $\xi_\sigma$ to arbitrary dimensional AdS gravity and check whether the relation $\xi_c=\xi_\sigma$ continues to hold.

In the next section, we compare and contrast the formalism used in this paper with the conventional theories of relativistic hydrodynamics. In particular, we would be interested in comparison with the conventional Israel-Stewart formalism.

\section{Israel-Stewart formalism}\label{sec:relhyd}
In this section, we give an extremely brief and non-exhaustive review of the conventional theories of relativistic hydrodynamics  \cite{Maartens:1996vi,Andersson:2006nr} and discuss how the work presented in this paper fits into that framework.

The first theories of relativistic viscous hydrodynamics are due to Eckart\cite{Eckart:1940te}, Landau and Lifshitz \cite{1959flme.book.....L}. These classical theories which are simple generalizations of their non-relativistic counterparts, assume a linear constitutive relation between the viscous stress $\pi^{\mu\nu}$ and the strain rate $\sigma^{\mu\nu}$. This linear approximation (called the Newtonian approximation) is the most familiar model in dissipative hydrodynamics and the fluids which obey such a relation are called Newtonian fluids.

Such a linear theory, however, leads to parabolic equations for the dissipative fluxes and predict very large speeds of propagation in situations with steep gradients, in contradiction with relativity and causality. It was noticed by many authors including Grad, Muller, Israel\cite{Israel:1976tn} and Stewart\cite{Israel:1979wp,1983AnPhy.151..466H} that one can easily solve this problem by including terms involving higher derivative corrections to the constitutive relations.\footnote{Many authors including Geroch\cite{Geroch:2001xs} have argued that the large speeds of propagation might not be a problem if the gradients required to produce them are so steep that they are beyond the domain of validity of hydrodynamics (We remind the reader that the hydrodynamics ceases to be valid if the ratio of mean free path to the length scale at consideration (i.e., the Knudsen number) is larger than one). But, this argument might not apply to all fluids - see \cite{Anile:1998hn,Herrera:2001ff} for further discussion of this issue. } The most simple extension is to add a non-zero relaxation time in the equation thus converting the problem into a hyperbolic system of equations. \footnote{If one is interested in rotational flows, one can further add other terms involving vorticity $\omega_{\mu\nu}$ and cross terms involving other hydrodynamic variables.} The resultant theory is called as causal  viscous hydrodynamics or Extended Irreversible Thermodynamics(EIT) or just Israel-Stewart theory.\footnote{Note that, there are other alternative solutions to the problem of causality in Newtonian hydrodynamics. One such class of models termed \textit{divergence type theories} were discussed by Geroch and Lindblom\cite{Geroch:1990bw} and under quite general conditions, these class of theories exhibit finite speeds of propagation\cite{Muller:1999in}. } 

This approach outlined above differs from the approach adopted here and elsewhere\cite{Bhattacharyya:2007jc,Baier:2007ix} in the holographic studies of $\mathcal{N}=4$ SYM. In particular, some of the terms appearing in the general derivative expansion of conformal fluids are absent in the conventional Israel-Stewart formalism\footnote{Further, the authors of the reference \cite{Baier:2007ix} argue that some of these terms would be absent even in a systematic derivation of Israel-Stewart formalism from Relativistic Kinetic theory via moment closures. }. 


One way of formulating Israel-Stewart theory is to consider dissipative fluxes like viscous stress and heat flow as new thermodynamic variables and treat entropy as a function of these new variables. In particular, one formulates the dynamics of such fluxes in a way that is consistent with the second law of thermodynamics. For a conformal fluid with no conserved charges, the viscoelastic stress in Israel-Stewart theory obeys an equation of the form\footnote{Note that often in the literature, $\tau_\omega$ is taken to be equal to $\tau_\pi$. We refrain from making such an identification here in order to facilitate easy comparison.}
\begin{equation}\label{israelstewart:eq}
\begin{split}
\pi^{\mu \nu} + \tau_\pi u^{\lambda}\mathcal{D}_{\lambda} \pi^{\mu\nu} &= - 2\ \eta\ \sigma^{\mu \nu} +\tau_\omega(\omega^{\mu}{}_\lambda\pi^{\lambda\nu}+\omega^{\nu}{}_\lambda\pi^{\lambda\mu})
\end{split}
\end{equation}
so that one can prove a version of the second law
\begin{equation}\label{isJs:eq}
\begin{split}
J^\lambda_s &= \left(s-\frac{\tau_\pi}{4\eta\tloc}\pi^{\mu\nu}\pi_{\mu\nu}\right) u^\lambda\\
\mathcal{D}_\lambda J^\lambda_s &= \frac{\pi^{\mu\nu}\pi_{\mu\nu}}{2\eta\tloc} \geq 0 \\
\end{split}
\end{equation}
There is now a wide literature devoted to the analysis of the equations above and this formalism has been recently applied to the phenomenology of heavy-ion collisions.\footnote{A non-exhaustive list of references include 
\cite{Muronga:2001zk,Muronga:2003ta,Heinz:2005bw,Baier:2006gy,Romatschke:2007jx,Song:2007ux,Bhalerao:2007ek} }

We can take the above equations and eliminate $\pi^{\mu\nu}$ in favor of $\sigma^{\mu\nu}$. We get the following expression which is exact up to second derivatives \begin{equation}\label{israelstewartmod:eq}
\begin{split}
\pi^{\mu \nu}  &= - 2\ \eta\brk{ \sigma^{\mu \nu} - \tau_\pi u^{\lambda}\mathcal{D}_{\lambda} \sigma^{\mu\nu} +\tau_\omega(\omega^{\mu}{}_\lambda\sigma^{\lambda\nu}+\omega^{\nu}{}_\lambda\sigma^{\lambda\mu})}\\\end{split}
\end{equation}
Comparing the equations so obtained with the equation\eqref{Tfinal:eq} , it is clear that an Israel-Stewart conformal fluid is a fluid with $\xi_\sigma,\xi_C$ and $\xi_\omega$ set to zero. Using the above expression, following the method employed in \S\ref{sec:entropy}, we can define an entropy current associated with this fluid obeying the second law.\footnote{Note however that the $J^\lambda_{s,diss}$ so obtained is the negative of what would be naively expected from equation\eqref{isJs:eq}. This apparent discrepancy can be traced to the ambiguity in the definition of $J^\lambda_{\text{S,diss}}$.}

However, as the previous sections make it clear, the Israel-Stewart conformal fluids form only a subset of conformal fluids. And more importantly, $\mathcal{N}=4$ SYM fluid lies outside the subset since it has $\xi_\sigma=\xi_C\neq 0$ . $\mathcal{N}=4$ SYM fluid has a shear-shear coupling (and a coupling to the Weyl curvature) which is absent in the conventional Israel Stewart formalism. Hence, the approach developed in the study of $\mathcal{N}=4$ SYM fluid should be looked upon as a generalization of the Israel Stewart formalism and the entropy current in the equation\eqref{jsfinal:eq} should be treated as a generalization of the Israel-Stewart expression in the equation\eqref{isJs:eq}. 

The main difference between the two formalisms lies in the way the viscoelastic stress is treated. As far as the contribution of the viscoelastic stress to the entropy current is concerned, Israel-Stewart formalism takes an extended thermodynamic point of view by assuming that all sources of viscoelastic stress contribute equally to the entropy current, whereas the entropy current proposed in this paper treats different sources of visco-elastic stress differently. Rather than assuming that the entropy density is solely a function of $\pi^{\mu\nu}$ , the entropy current is allowed to be a general function of the fluid velocity and its derivatives. Note that, despite going out of Israel Stewart formalism, we have succeeded in defining an entropy current which is consistent with the second law. \footnote{The author thanks Shiraz Minwalla for pointing out this distinction and for discussions about related issues.}

\section{Discussion and Conclusion}\label{sec:disc}
The holographic study of $\mathcal{N}=4$ SYM has already given us an interesting constitutive relation parametrised by \eqref{symparam:eq}. In this paper, an expression for the entropy current  that is consistent with this constitutive relation has been proposed via the simple requirement that the fluid in question should obey second law of thermodynamics. This gives a very specific expression for the entropy current of $\mathcal{N}=4$ SYM fluid. However, as has been mentioned before, demanding second law is often not sufficient to completely determine the entropy flux. A term with positive divergence can always be added to the entropy flux without violating second law. Given this fact, it is extremely important to have an independent holographic computation to check whether this proposal is indeed correct. 

We would like to remind the reader of an observation we made earlier - the rate of entropy production took a simpler form in the case of $\mathcal{N}=4$ SYM fluid. This is due to an interesting relation $\xi_\sigma=\xi_C$ which holds for $\mathcal{N}=4$ SYM fluid. It would be interesting to see whether this relation is an universal relation for conformal fluids with holographic duals in arbitrary dimensions.\footnote{The author would like to thank Dam Thanh Son for suggesting this possibility.} Further, it would be interesting to generalize the analysis of this paper to charged conformal fluids and find the corresponding entropy current.

We would like to note that indirectly, the expression given in this paper is also a proposal for an entropy current associated with the metric that is dual to the given fluid mechanical configuration. As of now, we do not have a very good prescription to calculate the entropy of such a metric configuration. This situation should be contrasted with the situation in the case of stationary black holes where the Bekenstein-Hawking entropy or more generally Wald entropy is believed to give a reliable prescription for calculating their entropy. Now that we have a specific proposal for the entropy current of a given metric configuration, a direct geometrical derivation of this entropy current would be a very interesting result.

In particular, the covariant formalism for conformal fluids that has been developed in this paper seems to be a natural setting in which the entropy current takes a simple form. Perhaps there exists a bulk interpretation of this formalism that provides the proper setting to look at the relation between the entropy and geometry. Given that the generalized second law in gravity is closely associated with the area increase theorem, it would be exciting to see how the area increase theorem in the bulk corresponds to the second law in the boundary. A detailed study of these issues may yield new insights regarding the relation between field theory and gravity.

On the other hand, in the gauge theory side, it would be interesting to compare the constitutive relation of the $\mathcal{N}=4$ SYM with that of the actual quark gluon plasma observed in RHIC. In particular, it would be interesting to work out  the effect of the new viscoelastic terms on the various observables of interest in heavy ion collisions like the elliptic flow\footnote{However, assuming that the second order effects are suppressed relative to the leading behaviour in the heavy ion collisions, it might be very difficult to extract any experimental signature of the viscoelastic behaviour. I would like to thank Paul Romatschke for pointing this out. }. The expression for the entropy current proposed here has an interesting structure which couples shear strain rate, expansion, acceleration and vorticity in a complicated way. A more thorough study of this expression might yield some insight on the entropy production and transport processes that happen at RHIC. It would be interesting to look at how the analysis in \cite{Dumitru:2007qr}, for example, would be changed, if we use the expression for the entropy current derived in this paper.\footnote{
Further, the expression here could be used, for example, to calculate and check the rate of entropy production in the numerical simulations of heavy-ion collisions (see \cite{Romatschke:2007jx,Song:2007ux} for some recent examples). The author wishes to thank Paul Romatschke for suggesting this possibility.}

%

\subsection*{Acknowledgements}
I would like to thank  Shiraz Minwalla for his advice, encouragement and support when this work was being done. I should thank Spenta Wadia, Saumen Datta and all the students in the TIFR theory students room - especially Sayantani Bhattacharya , Suvrat Raju, Basudeb Dasgupta and Suresh Nampuri for useful conversations. I thank Rajesh Gopakumar, Veronica Hubeny, Giuseppe Policastro, Mukund Rangamani, Paul Romatschke, Dam Thanh Son, Andrei Starinets and Misha Stephanov for their valuable comments.  I would also like to acknowledge useful discussions with Prerna Sharma regarding various models of viscoelasticity. Finally, I would like to acknowledge my debt to all those who have generously supported and encouraged the pursuit of science in India.

\section*{Appendices}
\appendix

\section{Some useful identities}\label{sec:identity}
In this appendix, we prove some identities that were used in the main body of this paper. In particular, we want to sketch the proof of the equations quoted in equation\eqref{theidentity:eq}.

First, we use the definition of $\mathcal{R}_{\mu\alpha\nu}{}^\lambda$ in terms of the commutator to write
\begin{equation}\label{basicid:eq}
\begin{split}
u^\alpha(\mathcal{R}_{\mu\alpha\nu}{}^\lambda u_\lambda + \mathcal{F}_{\mu\alpha} u_\nu ) &= -u^\alpha[\mathcal{D}_\mu,\mathcal{D}_\alpha]u_\nu 
\\
&=-\mathcal{D}_\mu(u^\alpha\mathcal{D}_\alpha u_\nu ) +(\mathcal{D}_\mu u^\alpha)(\mathcal{D}_\alpha u_\nu) + u^\alpha\mathcal{D}_\alpha(\mathcal{D}_\mu u_\nu )\\
&= \sigma_\mu{}^\alpha \sigma_{\alpha\nu} + \sigma_\mu{}^\alpha \omega_{\alpha\nu} - \sigma_\nu{}^\alpha \omega_{\alpha\mu} + \omega_\mu{}^\alpha \omega_{\alpha\nu} + u^\alpha\mathcal{D}_\alpha(\sigma_{\mu\nu}+\omega_{\mu\nu})\\
\end{split}
\end{equation}

Next, we multiply the expression above with $\sigma^{\mu\nu}$ and $\omega^{\mu\nu}$ respectively, and simplify the resulting expressions using the curvature identities in \S \ref{sec:curv} to get 
\begin{equation}\label{id1:eq}
\begin{split}
\sigma^{\mu\nu} C_{\mu\alpha\nu\beta} u^\alpha u^\beta -\sigma^{\mu\nu}\mathcal{S}_{\mu\nu} &= \sigma^{\mu\nu} \sigma_\mu{}^\alpha \sigma_{\alpha\nu} + \sigma^{\mu\nu} \omega_\mu{}^\alpha \omega_{\alpha\nu} + \sigma^{\mu\nu} u^\alpha\mathcal{D}_\alpha \sigma_{\mu\nu}\\
\frac{1}{2} \omega^{\mu\nu} \mathcal{F}_{\mu\nu} &= -2 \sigma_\mu{}^\alpha \omega_{\alpha\nu} \omega^{\nu\mu} + \omega^{\mu\nu} u^\alpha\mathcal{D}_\alpha \omega_{\mu\nu}\\
\end{split}
\end{equation}

The next step is to derive another identity which directly follows from the reduced Bianchi identity (See \eqref{redbianchi:eq} ) 
\begin{equation}\label{id2:eq}
\begin{split}
\mathcal{D}_\lambda\left[\frac{u_\mu(\mathcal{G}^{\mu\lambda}+\mathcal{F}^{\mu\lambda})}{d-2}\right]&= \frac{(\mathcal{D}_\lambda u_\mu)(\mathcal{G}^{\mu\lambda}+\mathcal{F}^{\mu\lambda})}{d-2}\\
&= \frac{(\mathcal{D}_\lambda u_\mu)(\mathcal{G}^{\mu\lambda}+\frac{d}{2}\mathcal{F}^{\mu\lambda}-\frac{d-2}{2}\mathcal{F}^{\mu\lambda})}{d-2} \\
&=\frac{\sigma_{\lambda\mu}(\mathcal{G}^{\mu\lambda}+\frac{d}{2}\mathcal{F}^{\mu\lambda})}{d-2}-\frac{1}{2}\omega_{\lambda\mu}\mathcal{F}^{\mu\lambda} \\
&=\sigma^{\mu\nu}\mathcal{S}_{\mu\nu}+\frac{1}{2}\omega_{\mu\nu}\mathcal{F}^{\mu\nu}\\
\end{split}
\end{equation}
where we have used the fact that $\mathcal{G}^{\mu\lambda}+\frac{d}{2}\mathcal{F}^{\mu\lambda}$ is a symmetric tensor.

We will need one more identity to finish the proof.
\begin{equation}\label{id3:eq}
\begin{split}
\mathcal{D}_\mu\brk{\frac{\mathcal{D}_\nu \omega^{\mu\nu}}{d-3}} &= \frac{1}{2(d-3)}[\mathcal{D}_\mu,\mathcal{D}_\nu] \omega^{\mu\nu}\\
& = \frac{3\mathcal{F}_{\mu\nu}\omega^{\mu\nu} 
+\mathcal{R}_{[\mu\nu]}\omega^{\mu\nu}}{2(d-3)} = -\frac{1}{2}\mathcal{F}_{\mu\nu}\omega^{\mu\nu} \\
\end{split}
\end{equation}

Using the above identities, it is now straightforward to get the equations quoted in \eqref{theidentity:eq}.
\begin{equation}
\begin{split}
\sigma^{\mu\nu} \omega_\mu{}^\alpha \omega_{\alpha\nu} &= \mathcal{D}_\lambda\brk{\frac{\omega^{\mu\nu}\omega_{\mu\nu}}{4}u^\lambda+
\frac{\mathcal{D}_\nu \omega^{\lambda\nu}}{2(d-3)}}\\
\sigma^{\mu\nu} \mathcal{C}_{\mu\alpha\nu\beta} u^\alpha u^\beta  &= \sigma^{\mu\nu} \sigma_\mu{}^\alpha \sigma_{\alpha\nu} + \mathcal{D}_\lambda\brk{ \frac{2\sigma^{\mu\nu}\sigma_{\mu\nu}+\omega^{\mu\nu}\omega_{\mu\nu}}{4} u^\lambda+\frac{u_\mu(\mathcal{G}^{\mu\lambda}+\mathcal{F}^{\mu\lambda})}{d-2}+
\frac{3\mathcal{D}_\nu \omega^{\lambda\nu}}{2(d-3)}}\\
\end{split}
\end{equation}

\section{Conformal Energy-Momentum tensor }\label{sec:cfnofR}
In this appendix, we list all the terms that can appear in the energy-momentum tensor of a conformal fluid and show that only a few of them are linearly independent.

In order to write down the most general derivative expansion of the viscoelastic stress $\pi^{\mu\nu}$, we list below all the Weyl- covariant second- rank tensors which are symmetric, transverse and traceless.
\begin{equation}
\begin{split}
&\quad \sigma^{\mu\nu}\ ,\quad   u^\lambda \mathcal{D}_\lambda \sigma^{\mu\nu}\ ,\quad [\omega^\mu{}_\lambda\sigma^{\lambda\nu}+\omega^\nu{}_\lambda\sigma^{\lambda\mu}] ,\\
&\quad [\sigma^\mu{}_\lambda\sigma^{\lambda\nu}-\frac{P^{\mu\nu}}{d-1}\sigma^{\alpha\beta}\sigma_{\alpha\beta}]\ ,\quad   [\omega^\mu{}_\lambda\omega^{\lambda\nu}+\frac{P^{\mu\nu}}{d-1}\omega^{\alpha\beta}\omega_{\alpha\beta}],\\
&\quad  C_{\mu\alpha\nu\beta}u^\alpha u^\beta, \quad [P^{\mu\lambda}P^{\nu\sigma}(\mathcal{R}_{\lambda\sigma}+\frac{d}{2}\mathcal{F}_{\lambda\sigma}) -\frac{P^{\mu\nu}}{d-1} P^{\lambda\sigma}\mathcal{R}_{\lambda\sigma} ], \\
&\quad  [P^{\mu\lambda}P^{\nu\sigma}(\mathcal{R}_{\lambda\alpha\sigma\beta}u^\alpha u^\beta -\frac{1}{2}\mathcal{F}_{\lambda\sigma})- \frac{P^{\mu\nu}}{d-1} P^{\lambda\sigma}\mathcal{R}_{\lambda\alpha\sigma\beta}u^\alpha u^\beta ]\\
\end{split}
\end{equation}
Note that, the different terms appearing above are not all independent . 

To show that we take the relation 
\begin{equation*}
-u^\alpha[\mathcal{D}_\mu,\mathcal{D}_\alpha] u_\nu = -u^\alpha\mathcal{D}_\mu\mathcal{D}_\alpha u_\nu + u^\alpha\mathcal{D}_\alpha\mathcal{D}_\mu u_\nu = (\mathcal{D}_\mu u^\alpha)(\mathcal{D}_\alpha u_\nu) + u^\alpha\mathcal{D}_\alpha(\mathcal{D}_\mu u_\nu)
\end{equation*}
and project out out the symmetric traceless transverse part to get
\begin{equation}\label{identity:eq}
\begin{split}
[P^{\mu\lambda}P^{\nu\sigma}(\mathcal{R}_{\lambda\alpha\sigma\beta}u^\alpha u^\beta &-\frac{1}{2}\mathcal{F}_{\lambda\sigma})- \frac{P^{\mu\nu}}{d-1} P^{\lambda\sigma}\mathcal{R}_{\lambda\alpha\sigma\beta}u^\alpha u^\beta ]\\
&= [\sigma^\mu{}_\lambda\sigma^{\lambda\nu}-\frac{P^{\mu\nu}}{d-1}\sigma^{\alpha\beta}\sigma_{\alpha\beta}]+ [\omega^\mu{}_\lambda\omega^{\lambda\nu}+\frac{P^{\mu\nu}}{d-1}\omega^{\alpha\beta}\omega_{\alpha\beta}]  + u^\lambda \mathcal{D}_\lambda \sigma^{\mu\nu}\\
\end{split}
\end{equation}

Further, if we denote by the subscript $TT$ the transverse traceless part, then we have using \eqref{weylschouten:eq}
\begin{equation*}
\begin{split}
[\mathcal{R}_{\lambda\sigma}+ (d-2) \mathcal{R}_{\lambda\alpha\sigma\beta}u^{\alpha}u^{\beta}]_{TT}&= [R_{\lambda\sigma}+ (d-2) R_{\lambda\alpha\sigma\beta}u^{\alpha}u^{\beta}]_{TT} =(d-2) C_{\lambda\alpha\sigma\beta}u^\alpha u^\beta \\
\end{split}
\end{equation*}

Hence, the independent terms that occur in a derivative expansion are 
\begin{equation}
\begin{split}
&\quad \sigma^{\mu\nu}\ ,\quad   u^\lambda \mathcal{D}_\lambda \sigma^{\mu\nu}\ ,\quad [\omega^\mu{}_\lambda\sigma^{\lambda\nu}+\omega^\nu{}_\lambda\sigma^{\lambda\mu}] ,\\
&\quad [\sigma^\mu{}_\lambda\sigma^{\lambda\nu}-\frac{P^{\mu\nu}}{d-1}\sigma^{\alpha\beta}\sigma_{\alpha\beta}]\ ,\quad   [\omega^\mu{}_\lambda\omega^{\lambda\nu}+\frac{P^{\mu\nu}}{d-1}\omega^{\alpha\beta}\omega_{\alpha\beta}],\\
&\quad  C_{\mu\alpha\nu\beta}u^\alpha u^\beta\\
\end{split}
\end{equation}
and so we obtain the derivative expansion in \eqref{derivexp:eq}.

\section{Notation}\label{app:notation}

We work in the $(-++\ldots)$ signature. $\mu,\nu$ denote space-time indices, $i,j=1 \ldots k$ label the $k$ different conserved charges. The dimensions of the spacetime in which the conformal fluid lives is denoted by $d$ .In the context of AdS/CFT, the dual AdS$_{d+1}$ space has $d+1$ spacetime dimensions. We use square brackets to denote antisymmetrisation. For example, $B_{[\mu\nu]}\equiv B_{\mu\nu}-B_{\nu\mu}$.

Our conventions for Christoffel symbols and the curvature tensors are fixed by the relations
\begin{equation}
\begin{split}
\nabla_{\mu}V^{\nu}&=\partial_{\mu}V^{\nu}+\Gamma_{\mu\lambda}{}^{\nu}V^{\lambda} \qquad \text{and}\qquad [\nabla_\mu,\nabla_\nu]V^\lambda=R_{\mu\nu\sigma}{}^{\lambda}V^\sigma .
\end{split}
\end{equation}

In the following table, the relevant equations are denoted by their respective equation numbers appearing inside parentheses.

\vp
  \centering
  \begin{tabular}{||r|l||r|l||}
    \hline
    Symbol & Definition & Symbol & Definition \\
    \hline
    $d$ & dimensions of spacetime & $p$ & Pressure \\
    $s$ & Proper entropy density & $\rho_i$ & Proper charge density  \\
    $\tloc$ & Fluid temperature & $\mu_i$ & Chemical potentials of the fluid \\
    $\nu_i$ & $\mu_i/\tloc$  & $\eta$ & Shear viscosity measured at \\
    $\tau_{\pi}$ & Stress relaxation time \eqref{Tfinal:eq} & & zero shear and vorticity \eqref{Tfinal:eq}\\
    $\tau_{\omega}$ & Shear vorticity coupling \eqref{Tfinal:eq}&$\xi_\sigma$ & Shear- shear coupling \eqref{Tfinal:eq} \\
    $\xi_C$ & Weyl Curvature coupling \eqref{Tfinal:eq} &$\xi_\omega$& Vorticity vorticity coupling \eqref{Tfinal:eq} \\
    \hline
%
    $T^{\mu\nu}$ & Energy-momentum tensor & $J^\mu_S$ & Entropy current \\
    $J^\mu_i$ & Charge currents & $u^\mu$ & Fluid velocity ($u^\mu u_\mu =-1$) \\
    $g_{\mu\nu}$ & Spacetime metric & $P^{\mu\nu}$ & Projection tensor, $g^{\mu\nu}+u^\mu u^\nu$ \\
    $a^\mu$ & Fluid acceleration \eqref{thetaaA:eq} &$\vartheta$&  Fluid expansion \eqref{thetaaA:eq} \\
    $\sigma_{\mu\nu}$ & Shear strain rate\eqref{Du:eq} & $\omega_{\mu\nu}$ & Fluid vorticity \eqref{Du:eq}\\
    $\pi_{\mu\nu}$ & Visco-elastic stress \eqref{dissdef:eq} &$\nu^\mu_{i}$ & Charge diffusion currents \eqref{dissdef:eq}\\
    \hline
%
     $\eta_i$ & Coefficients in & $\eta_0$ & $p/\tloc^d$ \\
     $ $ & derivative expansion\eqref{derivexp:eq} & $\eta_1$ & $-2\eta/\tloc^{d-1}$ \\      
     $\eta_2$ & $2\eta\tau_{\pi}/\tloc^{d-2}$ & & $\leq 0$ to satisfy second law \eqref{etacondn:eq}\\
     $\eta_3$ & $-2\eta\tau_{\omega}/\tloc^{d-2}$& $\eta_4$ & $\xi_\sigma/\tloc^{d-2}$ \\
      $\eta_5$& $\xi_\omega/\tloc^{d-2}$ & $\eta_6$ & -$\xi_C/\tloc^{d-2}$ \\
     \hline
%
     $\mathcal{D}_\mu$ & Weyl-covariant derivative \eqref{D:eq} & $\mathcal{A}_\mu$ & See  \eqref{thetaaA:eq}\\
     $\nabla_{\mu}$ & Lorentz-covariant derivative \eqref{gradu:eq} & $\Gamma_{\mu\nu}{}^\lambda$ & Christoffel connection  \\
     $R_{\mu\nu\lambda}{}^{\sigma}$ & Riemann Curvature \eqref{gradcomm:eq}& $\mathcal{R}_{\mu\nu\lambda}{}^{\sigma}$ & See \eqref{covcomm:eq}, \eqref{calcurv:eq} and \eqref{curvsym:eq}\\
     $\mathcal{F}_{\mu\nu}$ & $\nabla_\mu\mathcal{A}_\nu-\nabla_\nu\mathcal{A}_\mu$ & & \\
     $R_{\mu\nu},R$ & Ricci tensor/scalar & $\mathcal{R_{\mu\nu}},\mathcal{R}$& See \eqref{ricEin:eq} \\
     $G_{\mu\nu}$ & Einstein tensor & $\mathcal{G}_{\mu\nu}$ & See \eqref{ricEin:eq}\\
     $S_{\mu\nu},$ & Schouten tensor \eqref{schouten:eq}& $C_{\mu\nu\lambda\sigma},$ & Weyl Curvature \eqref{weylcurv:eq},\eqref{weylsym:eq}\\
      $\mathcal{S}_{\mu\nu}$ & & $\mathcal{C}_{\mu\nu\lambda\sigma}$&  \\
     \hline
\end{tabular}

\bibliographystyle{utcaps}
\bibliography{flumech}

\providecommand{\href}[2]{#2}\begingroup\raggedright\begin{thebibliography}{10}

\bibitem{Bhattacharyya:2007jc}
S.~Bhattacharyya, V.~E. Hubeny, S.~Minwalla, and M.~Rangamani, ``Nonlinear
  Fluid Dynamics from Gravity,''
\href{http://arXiv.org/abs/arXiv:0712.2456 [hep-th]}{{\tt arXiv:0712.2456
  [hep-th]}}.

\bibitem{Baier:2007ix}
R.~Baier, P.~Romatschke, D.~T. Son, A.~O. Starinets, and M.~A. Stephanov,
  ``Relativistic viscous hydrodynamics, conformal invariance, and holography,''
\href{http://arXiv.org/abs/arXiv:0712.2451 [hep-th]}{{\tt arXiv:0712.2451
  [hep-th]}}.

\bibitem{Rischke:1995ir}
D.~H. Rischke, S.~Bernard, and J.~A. Maruhn, ``Relativistic hydrodynamics for
  heavy ion collisions. 1. General aspects and expansion into vacuum,'' {\em
  Nucl. Phys.} {\bf A595} (1995) 346--382,
\href{http://arXiv.org/abs/nucl-th/9504018}{{\tt nucl-th/9504018}}.

\bibitem{Kolb:2003dz}
P.~F. Kolb and U.~W. Heinz, ``Hydrodynamic description of ultrarelativistic
  heavy-ion collisions,''
\href{http://arXiv.org/abs/nucl-th/0305084}{{\tt nucl-th/0305084}}.

\bibitem{Shuryak:2003xe}
E.~Shuryak, ``Why does the quark gluon plasma at RHIC behave as a nearly ideal
  fluid?,'' {\em Prog. Part. Nucl. Phys.} {\bf 53} (2004) 273--303,
\href{http://arXiv.org/abs/hep-ph/0312227}{{\tt hep-ph/0312227}}.

\bibitem{Adams:2005dq}
{\bf STAR} Collaboration, J.~Adams {\em et al.}, ``Experimental and theoretical
  challenges in the search for the quark gluon plasma: The STAR collaboration's
  critical assessment of the evidence from RHIC collisions,'' {\em Nucl. Phys.}
  {\bf A757} (2005) 102--183,
\href{http://arXiv.org/abs/nucl-ex/0501009}{{\tt nucl-ex/0501009}}.

\bibitem{Romatschke:2007mq}
P.~Romatschke and U.~Romatschke, ``Viscosity Information from Relativistic
  Nuclear Collisions: How Perfect is the Fluid Observed at RHIC?,'' {\em Phys.
  Rev. Lett.} {\bf 99} (2007) 172301,
\href{http://arXiv.org/abs/arXiv:0706.1522 [nucl-th]}{{\tt arXiv:0706.1522
  [nucl-th]}}.

\bibitem{Aharony:1999ti}
O.~Aharony, S.~S. Gubser, J.~M. Maldacena, H.~Ooguri, and Y.~Oz, ``Large N
  field theories, string theory and gravity,'' {\em Phys. Rept.} {\bf 323}
  (2000) 183--386,
\href{http://arXiv.org/abs/hep-th/9905111}{{\tt hep-th/9905111}}.

\bibitem{Klebanov:2000me}
I.~R. Klebanov, ``TASI lectures: Introduction to the AdS/CFT correspondence,''
\href{http://arXiv.org/abs/hep-th/0009139}{{\tt hep-th/0009139}}.

\bibitem{D'Hoker:2002aw}
E.~D'Hoker and D.~Z. Freedman, ``Supersymmetric gauge theories and the AdS/CFT
  correspondence,''
\href{http://arXiv.org/abs/hep-th/0201253}{{\tt hep-th/0201253}}.

\bibitem{Policastro:2002se}
G.~Policastro, D.~T. Son, and A.~O. Starinets, ``From AdS/CFT correspondence to
  hydrodynamics,'' {\em JHEP} {\bf 09} (2002) 043,
\href{http://arXiv.org/abs/hep-th/0205052}{{\tt hep-th/0205052}}.

\bibitem{Policastro:2002tn}
G.~Policastro, D.~T. Son, and A.~O. Starinets, ``From AdS/CFT correspondence to
  hydrodynamics. II: Sound waves,'' {\em JHEP} {\bf 12} (2002) 054,
\href{http://arXiv.org/abs/hep-th/0210220}{{\tt hep-th/0210220}}.

\bibitem{Herzog:2002fn}
C.~P. Herzog, ``The hydrodynamics of M-theory,'' {\em JHEP} {\bf 12} (2002)
  026,
\href{http://arXiv.org/abs/hep-th/0210126}{{\tt hep-th/0210126}}.

\bibitem{Kovtun:2003wp}
P.~Kovtun, D.~T. Son, and A.~O. Starinets, ``Holography and hydrodynamics:
  Diffusion on stretched horizons,'' {\em JHEP} {\bf 10} (2003) 064,
\href{http://arXiv.org/abs/hep-th/0309213}{{\tt hep-th/0309213}}.

\bibitem{Kovtun:2004de}
P.~Kovtun, D.~T. Son, and A.~O. Starinets, ``Viscosity in strongly interacting
  quantum field theories from black hole physics,'' {\em Phys. Rev. Lett.} {\bf
  94} (2005) 111601,
\href{http://arXiv.org/abs/hep-th/0405231}{{\tt hep-th/0405231}}.

\bibitem{Starinets:2005cy}
A.~O. Starinets, ``Transport coefficients of strongly coupled gauge theories:
  Insights from string theory,'' {\em Eur. Phys. J.} {\bf A29} (2006) 77--81,
\href{http://arXiv.org/abs/nucl-th/0511073}{{\tt nucl-th/0511073}}.

\bibitem{Benincasa:2006fu}
P.~Benincasa, A.~Buchel, and R.~Naryshkin, ``The shear viscosity of gauge
  theory plasma with chemical potentials,'' {\em Phys. Lett.} {\bf B645} (2007)
  309--313,
\href{http://arXiv.org/abs/hep-th/0610145}{{\tt hep-th/0610145}}.

\bibitem{Janik:2005zt}
R.~A. Janik and R.~Peschanski, ``Asymptotic perfect fluid dynamics as a
  consequence of AdS/CFT,'' {\em Phys. Rev.} {\bf D73} (2006) 045013,
\href{http://arXiv.org/abs/hep-th/0512162}{{\tt hep-th/0512162}}.

\bibitem{Janik:2006ft}
R.~A. Janik, ``Viscous plasma evolution from gravity using AdS/CFT,'' {\em
  Phys. Rev. Lett.} {\bf 98} (2007) 022302,
\href{http://arXiv.org/abs/hep-th/0610144}{{\tt hep-th/0610144}}.

\bibitem{Gubser:2006bz}
S.~S. Gubser, ``Drag force in AdS/CFT,'' {\em Phys. Rev.} {\bf D74} (2006)
  126005,
\href{http://arXiv.org/abs/hep-th/0605182}{{\tt hep-th/0605182}}.

\bibitem{Mas:2006dy}
J.~Mas, ``Shear viscosity from R-charged AdS black holes,'' {\em JHEP} {\bf 03}
  (2006) 016,
\href{http://arXiv.org/abs/hep-th/0601144}{{\tt hep-th/0601144}}.

\bibitem{Maeda:2006by}
K.~Maeda, M.~Natsuume, and T.~Okamura, ``Viscosity of gauge theory plasma with
  a chemical potential from AdS/CFT,'' {\em Phys. Rev.} {\bf D73} (2006)
  066013,
\href{http://arXiv.org/abs/hep-th/0602010}{{\tt hep-th/0602010}}.

\bibitem{Nakamura:2006ih}
S.~Nakamura and S.-J. Sin, ``A holographic dual of hydrodynamics,'' {\em JHEP}
  {\bf 09} (2006) 020,
\href{http://arXiv.org/abs/hep-th/0607123}{{\tt hep-th/0607123}}.

\bibitem{Saremi:2006ep}
O.~Saremi, ``The viscosity bound conjecture and hydrodynamics of M2- brane
  theory at finite chemical potential,'' {\em JHEP} {\bf 10} (2006) 083,
\href{http://arXiv.org/abs/hep-th/0601159}{{\tt hep-th/0601159}}.

\bibitem{Son:2006em}
D.~T. Son and A.~O. Starinets, ``Hydrodynamics of R-charged black holes,'' {\em
  JHEP} {\bf 03} (2006) 052,
\href{http://arXiv.org/abs/hep-th/0601157}{{\tt hep-th/0601157}}.

\bibitem{Lin:2006rf}
S.~Lin and E.~Shuryak, ``Toward the AdS/CFT gravity dual for High Energy
  Collisions: I.Falling into the AdS,''
\href{http://arXiv.org/abs/hep-ph/0610168}{{\tt hep-ph/0610168}}.

\bibitem{Lin:2007fa}
S.~Lin and E.~Shuryak, ``Toward the AdS/CFT Gravity Dual for High Energy
  Collisions: II. The Stress Tensor on the Boundary,''
\href{http://arXiv.org/abs/arXiv:0711.0736 [hep-th]}{{\tt arXiv:0711.0736
  [hep-th]}}.

\bibitem{Liu:2006nn}
H.~Liu, K.~Rajagopal, and U.~A. Wiedemann, ``An AdS/CFT calculation of
  screening in a hot wind,'' {\em Phys. Rev. Lett.} {\bf 98} (2007) 182301,
\href{http://arXiv.org/abs/hep-ph/0607062}{{\tt hep-ph/0607062}}.

\bibitem{Liu:2006ug}
H.~Liu, K.~Rajagopal, and U.~A. Wiedemann, ``Calculating the jet quenching
  parameter from AdS/CFT,'' {\em Phys. Rev. Lett.} {\bf 97} (2006) 182301,
\href{http://arXiv.org/abs/hep-ph/0605178}{{\tt hep-ph/0605178}}.

\bibitem{Heller:2007qt}
M.~P. Heller and R.~A. Janik, ``Viscous hydrodynamics relaxation time from
  AdS/CFT,'' {\em Phys. Rev.} {\bf D76} (2007) 025027,
\href{http://arXiv.org/abs/hep-th/0703243}{{\tt hep-th/0703243}}.

\bibitem{Kats:2007mq}
Y.~Kats and P.~Petrov, ``Effect of curvature squared corrections in AdS on the
  viscosity of the dual gauge theory,''
\href{http://arXiv.org/abs/arXiv:0712.0743 [hep-th]}{{\tt arXiv:0712.0743
  [hep-th]}}.

\bibitem{Kovchegov:2007pq}
Y.~V. Kovchegov and A.~Taliotis, ``Early time dynamics in heavy ion collisions
  from AdS/CFT correspondence,'' {\em Phys. Rev.} {\bf C76} (2007) 014905,
\href{http://arXiv.org/abs/arXiv:0705.1234 [hep-ph]}{{\tt arXiv:0705.1234
  [hep-ph]}}.

\bibitem{Myers:2007we}
R.~C. Myers, A.~O. Starinets, and R.~M. Thomson, ``Holographic spectral
  functions and diffusion constants for fundamental matter,'' {\em JHEP} {\bf
  11} (2007) 091,
\href{http://arXiv.org/abs/arXiv:0706.0162 [hep-th]}{{\tt arXiv:0706.0162
  [hep-th]}}.

\bibitem{Natsuume:2007ty}
M.~Natsuume and T.~Okamura, ``Causal hydrodynamics of gauge theory plasmas from
  AdS/CFT duality,''
\href{http://arXiv.org/abs/arXiv:0712.2916 [hep-th]}{{\tt arXiv:0712.2916
  [hep-th]}}.

\bibitem{Kajantie:2008rx}
K.~Kajantie, J.~Louko, and T.~Tahkokallio, ``Gravity dual of conformal matter
  collisions in 1+1 dimension,''
\href{http://arXiv.org/abs/arXiv:0801.0198 [hep-th]}{{\tt arXiv:0801.0198
  [hep-th]}}.

\bibitem{1984ucp..book.....W}
R.~M. {Wald}, {\em {General relativity}}.
\newblock Chicago, University of Chicago Press, 1984, 504 p., 1984.

\bibitem{1992JMP....33.2633H}
G.~S. {Hall}, ``{Weyl manifolds and connections},'' {\em Journal of
  Mathematical Physics} {\bf 33} (July, 1992) 2633--2638.

\bibitem{1959flme.book.....L}
L.~D. {Landau} and E.~M. {Lifshitz}, {\em {Fluid mechanics}}.
\newblock Course of theoretical physics, Oxford: Pergamon Press, 1959.

\bibitem{Maartens:1996vi}
R.~Maartens, ``Causal thermodynamics in relativity,''
\href{http://arXiv.org/abs/astro-ph/9609119}{{\tt astro-ph/9609119}}.

\bibitem{Andersson:2006nr}
N.~Andersson and G.~L. Comer, ``Relativistic fluid dynamics: Physics for many
  different scales,''
\href{http://arXiv.org/abs/gr-qc/0605010}{{\tt gr-qc/0605010}}.

\bibitem{Eckart:1940te}
C.~Eckart, ``The Thermodynamics of irreversible processes. 3. Relativistic
  theory of the simple fluid,'' {\em Phys. Rev.} {\bf 58} (1940)
919--924.

\bibitem{Israel:1976tn}
W.~Israel, ``Nonstationary irreversible thermodynamics: A Causal relativistic
  theory,'' {\em Ann. Phys.} {\bf 100} (1976)
310--331.

\bibitem{Israel:1979wp}
W.~Israel and J.~M. Stewart, ``Transient relativistic thermodynamics and
  kinetic theory,'' {\em Ann. Phys.} {\bf 118} (1979)
341--372.

\bibitem{1983AnPhy.151..466H}
W.~A. {Hiscock} and L.~{Lindblom}, ``{Stability and causality in dissipative
  relativistic fluids.},'' {\em Annals of Physics} {\bf 151} (1983) 466--496.

\bibitem{Geroch:2001xs}
R.~Geroch, ``On Hyperbolic ''Theories'' of Relativistic Dissipative Fluids,''
\href{http://arXiv.org/abs/gr-qc/0103112}{{\tt gr-qc/0103112}}.

\bibitem{Anile:1998hn}
A.~M. Anile, D.~Pavon, and V.~Romano, ``The case for hyperbolic theories of
  dissipation in relativistic fluids,''
\href{http://arXiv.org/abs/gr-qc/9810014}{{\tt gr-qc/9810014}}.

\bibitem{Herrera:2001ff}
L.~Herrera and D.~Pavon, ``Hyperbolic theories of dissipation: Why and when do
  we need them?,'' {\em Physica} {\bf A307} (2002) 121--130,
\href{http://arXiv.org/abs/gr-qc/0111112}{{\tt gr-qc/0111112}}.

\bibitem{Geroch:1990bw}
R.~Geroch and L.~Lindblom, ``Dissipative relativistic fluid theories of
  divergence type,'' {\em Phys. Rev.} {\bf D41} (1990)
1855.

\bibitem{Muller:1999in}
I.~Muller, ``Speeds of propagation in classical and relativistic extended
  thermodynamics,'' {\em Living Rev. Rel.} {\bf 2} (1999)
1.

\bibitem{Muronga:2001zk}
A.~Muronga, ``Second order dissipative fluid dynamics for ultra- relativistic
  nuclear collisions,'' {\em Phys. Rev. Lett.} {\bf 88} (2002) 062302,
\href{http://arXiv.org/abs/nucl-th/0104064}{{\tt nucl-th/0104064}}.

\bibitem{Muronga:2003ta}
A.~Muronga, ``Causal Theories of Dissipative Relativistic Fluid Dynamics for
  Nuclear Collisions,'' {\em Phys. Rev.} {\bf C69} (2004) 034903,
\href{http://arXiv.org/abs/nucl-th/0309055}{{\tt nucl-th/0309055}}.

\bibitem{Heinz:2005bw}
U.~W. Heinz, H.~Song, and A.~K. Chaudhuri, ``Dissipative hydrodynamics for
  viscous relativistic fluids,'' {\em Phys. Rev.} {\bf C73} (2006) 034904,
\href{http://arXiv.org/abs/nucl-th/0510014}{{\tt nucl-th/0510014}}.

\bibitem{Baier:2006gy}
R.~Baier and P.~Romatschke, ``Causal viscous hydrodynamics for central
  heavy-ion collisions,'' {\em Eur. Phys. J.} {\bf C51} (2007) 677--687,
\href{http://arXiv.org/abs/nucl-th/0610108}{{\tt nucl-th/0610108}}.

\bibitem{Romatschke:2007jx}
P.~Romatschke, ``Causal viscous hydrodynamics for central heavy-ion collisions.
  II: Meson spectra and HBT radii,'' {\em Eur. Phys. J.} {\bf C52} (2007)
  203--209,
\href{http://arXiv.org/abs/nucl-th/0701032}{{\tt nucl-th/0701032}}.

\bibitem{Song:2007ux}
H.~Song and U.~W. Heinz, ``Causal viscous hydrodynamics in 2+1 dimensions for
  relativistic heavy-ion collisions,''
\href{http://arXiv.org/abs/arXiv:0712.3715 [nucl-th]}{{\tt arXiv:0712.3715
  [nucl-th]}}.

\bibitem{Bhalerao:2007ek}
R.~S. Bhalerao and S.~Gupta, ``Aspects of causal viscous hydrodynamics,''
\href{http://arXiv.org/abs/arXiv:0706.3428 [nucl-th]}{{\tt arXiv:0706.3428
  [nucl-th]}}.

\bibitem{Dumitru:2007qr}
A.~Dumitru, E.~Molnar, and Y.~Nara, ``Entropy production in high-energy
  heavy-ion collisions and the correlation of shear viscosity and
  thermalization time,'' {\em Phys. Rev.} {\bf C76} (2007) 024910,
\href{http://arXiv.org/abs/arXiv:0706.2203 [nucl-th]}{{\tt arXiv:0706.2203
  [nucl-th]}}.

\end{thebibliography}\endgroup

\end{document}